\documentclass[preprint,aps]{revtex4}

\begin{document}

\title{Comprehensive Solution to the Cosmological Constant, Zero-Point Energy, and Quantum Gravity Problems}

\author{Philip~D.~Mannheim}

\affiliation{Department of Physics\\ University of Connecticut\\ Storrs, CT
06269, USA
\\ {\tt philip.mannheim@uconn.edu}}

\date{August 19, 2010}

\begin{abstract}
We present a solution to the cosmological constant, the zero-point energy, and the quantum gravity problems within a single comprehensive framework. We show that in quantum theories of gravity in which the zero-point energy density of the gravitational field is well-defined, the cosmological constant and zero-point energy problems solve each other by mutual cancellation between the cosmological constant and the matter and gravitational field zero-point energy densities. Because of this cancellation, regulation of the matter field zero-point energy density is not needed, and thus does not cause any trace anomaly to arise. We exhibit our results in two theories of gravity that are well-defined quantum-mechanically. Both of these theories are locally conformal invariant, quantum Einstein gravity in two dimensions and Weyl-tensor-based quantum conformal gravity in four dimensions (a fourth-order derivative quantum theory of the type that Bender and Mannheim have recently shown to be ghost-free and unitary). Central to our approach is the requirement that any and all departures of the geometry from Minkowski are to be brought about by quantum mechanics alone. Consequently, there have to be no fundamental classical fields, and all mass scales have to be generated by dynamical condensates. In such a situation the trace of the matter field energy-momentum tensor is zero, a constraint that obliges its cosmological constant and zero-point contributions to cancel each other identically, no matter how large they might be. In our approach quantization of the gravitational field is caused by its coupling to quantized matter fields, with the gravitational field not needing any independent quantization of its own. With there being no a priori classical curvature, one does not have to make it compatible with quantization.

\end{abstract}

\maketitle

\section{Vacuum Energy Problem and the Need for Quantum Gravity}
\label{S1}

Despite the fact that general relativity intricately connects the geometrical Einstein tensor to the matter field energy-momentum tensor, in the theory the matter field energy-momentum tensor can be independently assigned as it is not constrained to be zero in the flat spacetime limit where gravity is absent. With a fundamental-physics-based matter field energy-momentum tensor already possessing both cosmological constant and zero-point fluctuation contributions in flat spacetime, standard gravity has no control over them, and has to abide with whatever values flat space physics bequeaths to it. To give gravity control of these contributions then, one should consider seeking an approach to gravity in which curvature is tied into the matter field energy-momentum tensor from the outset, with the matter field energy-momentum tensor and the gravitational tensor both being required to be zero in each other's absence. To achieve this we therefore posit that on the matter field side there be no fundamental c-number classical fields in nature and that the matter field energy-momentum tensor be completely quantum-mechanical. (Except for the presence of a fundamental scalar Higgs field this is the standard $SU(3)\times SU(2)\times U(1)$ picture, and so we only need to modify the standard model of particle physics  by replacing fundamental Higgs fields by dynamical fermion condensates.) With such an energy-momentum tensor as its source, we equally posit that non-zero curvature be due to quantum mechanics alone, with spacetime then being Minkowski (or conformally equivalent to it) in the absence of quantum mechanics. The role of quantization is thus to make both the gravitational and matter field components of the gravitational equations of motion be non-zero simultaneously, with the gravitational field being expanded as a power series in Planck's constant rather than as a power series in the gravitational coupling constant. As we will see, this will lead us to a theory of gravity in which the zero-point fluctuations of the gravitational and matter fields and the contribution of a condensate-induced cosmological constant term all mutually cancel each other identically no matter how big they might be.

In order to be able to implement these ideas at all, we will need a theory of quantum gravity that is consistent in four spacetime dimensions (4D). And with standard 4D Einstein gravity not being renormalizable, and with the ghost/unitarity problem associated with fourth-order derivative theories having recently been resolved \cite{R1,R2}, we are led to consider conformal gravity, a fourth-order derivative theory of gravity that is renormalizable in 4D.  In a sense this is actually a natural choice since when all fermion and gauge boson masses are generated dynamically, the standard $SU(3)\times SU(2)\times U(1)$ model of particle physics is conformally invariant at the level of the Lagrangian, and one should thus expect that whatever is to be the gravitational tensor is to have the same symmetry as the source to which it is to couple. In our study we shall find that the tracelessness property that a conformal invariant energy-momentum tensor has to possess will play a crucial role, as it forces all of its various zero-point and cosmological constant components to exactly cancel each other in the vanishing trace no matter how large they might actually be. 

To illustrate how our ideas are to work, it is instructive to consider a conformal invariant theory that is simpler to treat than fourth-order derivative 4D conformal gravity, and so first we study Einstein gravity in two dimensions (2D), as this is the spacetime dimension in which standard gravity is conformal invariant. Now classically the 2D Einstein-Hilbert action is a total divergence, and so 2D classical Einstein gravity does not exist. Moreover, since the classical action is a total divergence, one cannot construct canonical conjugates via functional differentiation and the theory cannot be quantized canonically. Nonetheless, if one maintains the ordering in which the metric components and their derivatives appear in the action, in evaluating $(-g)^{1/2}R^{\alpha}_{\phantom{\alpha}\alpha}$ one finds that in addition to total derivatives, one also encounters various commutator terms (generically, $A\partial_{\mu}B+B\partial_{\mu}A =\partial_{\mu}(AB)+[B,\partial_{\mu}A]$.) While these specific commutator terms are not in the form of conventional canonical commutator terms,  their non-vanishing (as is enforced by coupling to a quantum-mechanical fermion) will enable the quantum 2D Einstein theory to be non-trivial. As we shall see, this non-triviality will lead to a gravitational field zero-point energy density that will precisely cancel that of a conformal invariant fermion matter field source. 

Even though quantum 2D Einstein gravity has a non-vanishing zero-point energy sector, that is essentially all that there is to the theory. Specifically, since the 2D Einstein-Hilbert classical action is a total divergence in all classical paths, be they stationary or non-stationary, the path integral is just a constant and there is no quantum scattering. However, as constructed in field theory, the path integral does not give matrix elements of the products of the quantum field operator themselves, but only of normal ordered products of them. The path integral does not give the energy of the vacuum but only the energy after the infinite zero-point contribution has been subtracted out (as must be the case if the $e^{-E_0\tau}$ deep Euclidean time limit of the path integral is to be finite and the path integral is to exist). Now while path integral quantization suffices for flat spacetime where only energy differences are measurable, for gravity one additionally needs to know where the zero of energy actually is as gravity couples to energy itself rather than to energy difference. Thus for gravity one needs some information that is not contained in the path integral, namely zero-point information, with zero-point energy being able to exist and be non-zero even if the path integral itself is trivial. Thus beyond our own specific interest here in providing an explicitly solvable model in which the zero-point contributions of the matter fields and the gravitational field identically cancel each other, 2D quantum Einstein gravity is also of interest in that it shows that the zero-point energy is on a different footing than all of the other aspects of quantum field theory. Identifying the specific status in quantum field theory that the zero-point energy thus has is one of the central elements of the solution to  the zero-point energy problem that we present here.

The present paper is organized into two main blocks. In Sec. \ref{U1} and its four subsections (\ref{S2}, \ref{S3}, \ref{S4}, \ref{S5}) we discuss the two-dimensional case, and in Sec. \ref{U2} and its three subsections (\ref{S6}, \ref{S7}, \ref{S8}) we discuss the four-dimensional case. In Sec. \ref{S2} we discuss the general structure of 2D quantum Einstein gravity and in Sec. \ref{S3} show how the fermion and gravitational zero-point energy densities cancel each other when the fermion is massless and non-self-interacting. In Sec. \ref{S4} we identify some general aspects of this analysis, and show how by virtue of being infinite, the matter field zero-point energy evades the well-known theorem that the energy of a Lorentz invariant vacuum has to be zero. In Sec. \ref{S5}  we extend the analysis of 2D quantum gravity to include 4-fermion interactions (interactions that are both renormalizable and conformal invariant in 2D), and show that the zero-point contributions and cosmological constant terms all cancel each other after the 4-fermion theory is spontaneously broken via the self-consistent Nambu-Jona-Lasinio mechanism \cite{R3}. In Sec. \ref{S6} we show how our ideas generalize to 4D conformal gravity, and in Secs. \ref{S7} and \ref{S8} we present the resolution of the ghost problem of fourth-order derivative theories via ${\cal P}{\cal T}$ symmetry given in \cite{R1,R2}, and discuss its relevance to the vacuum energy problem. In addition in Sec. \ref{S7} we establish that, just like 4D conformal gravity, 2D quantum Einstein gravity  is also a ${\cal P}{\cal T}$ theory. Finally, in  Sec. \ref{S9} we present our conclusions. Since this paper is somewhat long, for the benefit of the reader  in \cite{R3p} we have provided a short version of our manuscript that concentrates on the two-dimensional case.

\section{The Two-Dimensional Case}
\label{U1}

\subsection{Non-Triviality of 2D Quantum Einstein Gravity}
\label{S2}

Since the issue for 2D quantum Einstein gravity is the ordering of the fields, we shall adopt the convention of defining the Riemann tensor once and for all according to the ordering
\begin{equation}
R_{\lambda\mu\nu\kappa}=\frac{1}{2}\left(\partial_{\kappa}\partial_{\mu}
g_{\lambda\nu}  -\partial_{\kappa}\partial_{\lambda} g_{\mu\nu}  
-\partial_{\nu}\partial_{\mu}g_{\lambda\kappa} 
+\partial_{\nu}\partial_{\lambda}g_{\mu\kappa})
+g_{\eta\sigma}(\Gamma^{\eta}_{\nu\lambda}\Gamma^{\sigma}_{\mu\kappa}
-\Gamma^{\eta}_{\kappa\lambda}\Gamma^{\sigma}_{\mu\nu}\right),
\label{E1}
\end{equation}
where $\Gamma^{\alpha}_{\mu\kappa}$ is ordered according to $\Gamma^{\alpha}_{\mu\kappa}=(1/2)g^{\alpha\beta}(\partial_{\mu}g_{\beta\kappa}+\partial_{\kappa}g_{\beta\mu}-\partial_{\beta}g_{\mu\kappa})$, and shall provisionally define the Ricci tensor, the Ricci scalar and the Einstein tensor according to the ordering $R_{\mu\kappa}=g^{\nu\lambda}R_{\lambda\mu\nu\kappa}$, $R^{\alpha}_{\phantom{\alpha}\alpha}=g^{\mu\kappa}R_{\mu\kappa}$ and $G_{\mu\kappa}=R_{\mu\kappa} -(1/2)g_{\mu\kappa}R^{\alpha}_{\phantom{\alpha}\alpha}$. If we perturb to second order around flat spacetime according to $g_{\mu\nu}=\eta_{\mu\nu}+h_{\mu\nu}$, $g^{\mu\nu}=\eta^{\mu\nu}-h^{\mu\nu}+h^{\mu}_{\phantom{\mu}\sigma}h^{\sigma\nu}$, $(-g)^{1/2}=1+h/2+h^2/8-h_{\mu\nu}h^{\mu\nu}/4$ where $h=\eta^{\mu\nu}h_{\mu\nu}$, then in 2D we find that to first order in $h_{\mu\nu}$ the first order Einstein tensor $G_{\mu\nu}(1)$ vanishes kinematically (as it of course must since there is no ordering issue to first order and the 2D $G_{\mu\nu}(1)$ already vanishes classically). However, in second order we find that $G_{\mu\nu}(2)$ evaluates to the explicitly non-zero
\begin{eqnarray}
G_{00}(2)&=&\frac{1}{4}[\partial_0h_{00},\partial_1h_{01}]+\frac{1}{4}[\partial_1h_{11},\partial_0h_{01}] +\frac{1}{8}[\partial_0h_{11},\partial_0h_{00}] +\frac{1}{8}[\partial_1h_{00},\partial_1h_{11}]=G_{11}(2)
\nonumber \\
&=&\frac{1}{4}[\partial_0h_{00},\partial_{\mu}h^{\mu}_{\phantom{\mu}0}]-\frac{1}{4}[\partial_1h_{11},\partial_{\mu}h^{\mu}_{\phantom{\mu}1}] -\frac{1}{8}[\partial_0h_{11},\partial_0h] +\frac{1}{8}[\partial_1h_{00},\partial_1h],
\nonumber\\
G_{01}(2)&=&\frac{1}{4}[\partial_1h_{00},\partial_0h_{11}]+\frac{1}{4}[\partial_0h_{00},\partial_1h_{00}] +\frac{1}{2}[\partial_0h_{11},\partial_0h_{01}],
\nonumber\\
G_{10}(2)&=&\frac{1}{4}[\partial_1h_{11},\partial_0h_{11}]-\frac{1}{4}[\partial_1h_{00},\partial_0h_{11}]+ \frac{1}{2}[\partial_1h_{00},\partial_1h_{01}],
\nonumber\\
\frac{1}{2}G_{01}(2)&+&\frac{1}{2}G_{10}(2)
\nonumber\\
&=&\frac{1}{8}[\partial_0h_{00},\partial_1h]
-\frac{1}{8}[\partial_1h_{11},\partial_0h]
-\frac{1}{4}[\partial_0h_{11},\partial_{\mu}h^{\mu}_{\phantom{\mu}1}]
+ \frac{1}{4}[\partial_1h_{00},\partial_{\mu}h^{\mu}_{\phantom{\mu}0}],
\label{E2}
\end{eqnarray}
where for a metric with signature $\eta_{\mu\nu}={\rm diag}(-1,1)$, $\partial_{\mu}h^{\mu}_{\phantom{\mu}0}=-\partial_0h_{00}+\partial_1h_{01}$, $\partial_{\mu}h^{\mu}_{\phantom{\mu}1}=-\partial_0h_{01}+\partial_1h_{11}$, and $h=\eta^{\mu\nu}h_{\mu\nu}=-h_{00}+h_{11}$.
As is to be expected, the 2D $G_{\mu\nu}(2)$ is purely in the form of commutator terms since it has to vanish identically in the classical limit. 

Now while a classical quantity may obey various identities, some of these identities could fail quantum-mechanically when ordering is taken into account, and thus need to be considered anew. From (\ref{E2}) we see that the tracelessness property $\eta^{\mu\nu}G_{\mu\nu}(2)=0$  of a conformal invariant theory is preserved. However, the quantum $G^{\mu\nu}(2)$ is seen not to be symmetric in its indices. In the presence of ordering the quantity $g_{\eta\sigma}(\Gamma^{\eta}_{\nu\lambda}\Gamma^{\sigma}_{\mu\kappa}
-\Gamma^{\eta}_{\kappa\lambda}\Gamma^{\sigma}_{\mu\nu})$ that appears in  (\ref{E1}) is not the same as $g_{\eta\sigma}(\Gamma^{\sigma}_{\mu\kappa}\Gamma^{\eta}_{\nu\lambda}
-\Gamma^{\sigma}_{\mu\nu}\Gamma^{\eta}_{\kappa\lambda})$, and $R_{\mu\kappa}=g^{\nu\lambda}R_{\lambda\mu\nu\kappa}$ is not the same as $R_{\kappa\mu}=g^{\nu\lambda}R_{\lambda\kappa\nu\mu}$, with it actually being because of the ordering properties of products of Christoffel symbols that the quantum $G_{\mu\nu}(2)$ is non-zero in the first place. To rectify this lack of symmetry, in the following we shall redefine $R^{\mu\nu}$ to denote the symmetrized form $R^{\mu\nu}/2+R^{\nu\mu}/2$ (and thus redefine $G^{\mu\nu}$ to denote the symmetrized form $G^{\mu\nu}/2+G^{\nu\mu}/2$), a redefinition that has no consequences classically. However, even with this rectification, the quantum $G^{\mu\nu}(2)$ is found not to obey the Bianchi identity $\partial_{\mu} G^{\mu\nu}(2)=0$ that it would obey in the classical case. Specifically, the components of the rectified 2D $\partial_{\mu} G^{\mu\nu}(2)$ are found to evaluate to 
\begin{eqnarray}
\partial_{\mu} G^{\mu}_{\phantom{\mu} 0}(2)&=&\frac{1}{4}[\nabla^2h_{00},\partial_1h_{01}]-\frac{1}{4}[\nabla^2h_{01},\partial_1h_{11}]
-\frac{1}{4}[\partial_0\partial_1h_{01},\partial_0h]-\frac{1}{8}[\partial_0\partial_1(h_{00}+h_{11}),\partial_1h]
\nonumber\\
&&+\frac{1}{4}[\partial_1^2h_{01},\partial_1h]
+\frac{1}{8}[\nabla^2h,\partial_0h_{00}]
+\frac{1}{8}[\partial_1^2h_{11},\partial_0h]
+\frac{1}{8}[\partial_0^2h_{00},\partial_0h],
\nonumber\\
\partial_{\mu} G^{\mu}_{\phantom{\mu} 1}(2)&=&\frac{1}{4}[\nabla^2h_{11},\partial_0h_{01}]-\frac{1}{4}[\nabla^2h_{01},\partial_0h_{00}]
-\frac{1}{4}[\partial_0\partial_1h_{01},\partial_1h]-\frac{1}{8}[\partial_0\partial_1(h_{00}+h_{11}),\partial_0h]
\nonumber\\
&&+\frac{1}{4}[\partial_0^2h_{01},\partial_0h]
-\frac{1}{8}[\nabla^2h,\partial_1h_{11}]
+\frac{1}{8}[\partial_0^2h_{00},\partial_1h]
+\frac{1}{8}[\partial_1^2h_{11},\partial_1h],
\label{E3}
\end{eqnarray}
where $\nabla^2=-\partial_0^2+\partial_1^2$. However, since there are no first order constraints on the 2D $G^{\mu\nu}(1)$, there is nothing to force $ \partial_{\mu} G^{\mu}_{\phantom{\mu} \nu}(2)$ to vanish.

To address this issue we introduce a free massless fermion, to give a total conformal invariant action of the form $I=I_{\rm GRAV}+I_{\rm M}$, where the gravitational and matter field actions are given by
\begin{eqnarray} 
I_{\rm GRAV}&=&-\frac{1}{2\kappa_2^2}\int d^2x (-g)^{1/2}R^{\alpha}_{\phantom{\alpha}\alpha}, 
\nonumber\\
I_{\rm M}&=&-\frac{1}{2}\int d^2x (-g)^{1/2}\left[i\hbar\bar{\psi}\gamma^{\mu}(x)[\partial_\mu+\Gamma_\mu(x)] \psi -i\hbar\bar{\psi}[\stackrel{\leftarrow}{\partial_\mu}+\Gamma_{\mu}(x)]\gamma^{\mu}(x)\psi\right], 
\label{E4}
\end{eqnarray}
and where the $\gamma^{\mu}(x)$ are the vierbein-dependent Dirac gamma matrices, $\Gamma_{\mu}(x)$ is the fermion spin connection, and $\bar{\psi}=\psi^{\dagger}\gamma^0$. For the fermion sector we construct a matter energy-momentum tensor $T_{\rm M}^{\mu\nu}=2(\delta I_{\rm M}/\delta g_{\mu\nu})/(-g)^{1/2}$ of the form
\begin{eqnarray}
T_{\rm M}^{\mu\nu}&=&\frac{i\hbar}{4} \bar{\psi} \gamma^{\mu}(x)[\partial^{\nu}+\Gamma^\nu(x)] \psi  +\frac{i\hbar}{4} \bar{\psi} \gamma^{\nu}(x)[\partial^{\mu}+\Gamma^\mu(x)] \psi +{\rm H.~c.}
\nonumber \\
&&
-\frac{1}{2}g_{\mu\nu}\left[i\hbar\bar{\psi}\gamma^{\alpha}(x)[\partial_\alpha+\Gamma_\alpha(x)] \psi -i\hbar\bar{\psi}[\stackrel{\leftarrow}{\partial_\alpha}+\Gamma_{\alpha}(x)]\gamma^{\alpha}(x)\psi\right],
\label{E5}
\end{eqnarray}
with the gravitational equation of motion taking the form
\begin{equation}
\frac{1}{\kappa_2^2}G^{\mu\nu}+T^{\mu\nu}_{\rm M}=0.
\label{E6}
\end{equation}
In solutions to the fermion equation of motion $i\hbar\gamma^{\mu}(x)[\partial_\mu+\Gamma_\mu(x)] \psi =0$, we find that $T^{\mu\nu}_{\rm M}$ is both traceless and covariantly conserved. Now as constructed, one could take the trivial $\psi=0$ solution to be the solution to the Dirac equation, with $T^{\mu\nu}_{\rm M}$, and thus $G^{\mu\nu}$, then both being zero identically. However, if we quantize the fermion according to the standard anticommutation relation $\{\psi_{\alpha}({ x},t),\psi^{\dagger}_{\beta}({ x}^{\prime},t)\}=\delta({ x}-{ x}^{\prime})\delta_{\alpha,\beta}$, this will not only force $T^{\mu\nu}_{\rm M}$ to be non-zero, through (\ref{E6}) $G^{\mu\nu}$ will be forced to be non-zero as well. Quantization of the matter field thus forces quantization of the gravitational field, and thus forces the quantum 2D $G^{\mu\nu}$ to be non-zero even though the classical 2D $G^{\mu\nu}$ vanishes identically. Thus even though 2D quantum gravity has no classical limit and cannot be quantized canonically either starting from classical Poisson brackets or via path integral quantization, the theory nonetheless still exists as a non-empty quantum theory. 

Now we note that the conservation of the matter field $T^{\mu\nu}_{\rm M}$ is secured via the matter field equations of motion alone without any need to utilize the gravitational equations of motion at all. Consequently,  when (\ref{E6}) is imposed, it follows from it that the quantum $G^{\mu\nu}$ must be covariantly conserved too. We thus see a total reversal of the familiar classical gravity situation. In the classical case the Einstein tensor is kinematically conserved (the Bianchi identity), a conservation that holds in every classical variational gravitational path be it stationary or non-stationary, with the Einstein equations then forcing the conservation of $T^{\mu\nu}_{\rm M}$ because of the structure of the gravity sector. In the quantum case there is no kinematic Bianchi identity and one instead needs to start from the conservation of $T^{\mu\nu}_{\rm M}$ associated with the structure of the matter sector, and then use the gravitational equations of motion to obtain the conservation of  $G^{\mu\nu}$, with its conservation only holding on the stationary gravitational path that actually satisfies the gravitational field equations. 

As regards geodesic behavior, it too can be derived without needing to appeal to the Bianchi identity. Specifically, if one starts with the wave equation obeyed by the matter field in some external gravitational background, in the short wavelength limit, the wave function eikonalizes, with the rays normal to the wavefronts then being geodesic (see e.g. \cite{R4}). Thus again one can bypass the usual Bianchi-identity-based discussion.

For the case when gravity is quantum-mechanically linearized around classical flat spacetime so that the fluctuation $h_{\mu\nu}$ is to then be of order $\hbar^{1/2}$, the term that is second order in $h_{\mu\nu}$ in $(1/\kappa_2^2)G^{\mu\nu}$  is balanced by the term in $T^{\mu\nu}_{\rm M}$ that is associated with the quantization of a free fermion in a classical flat spacetime as both terms are of order $\hbar$, with curvature corrections to $T^{\mu\nu}_{\rm M}$ being higher order in $h_{\mu\nu}$. To second order in $h_{\mu\nu}$ then, the covariant conservation of $G^{\mu\nu}(2)$ as now enforced in (\ref{E3}) by the coupling to the fermion sector then entails that the quantized gravitational field components can be taken to obey the equations 
\begin{equation}
\nabla^2 h_{00}=0,\qquad \nabla^2 h_{01}=0,\qquad \nabla^2 h_{11}=0,\qquad h=-h_{00}+h_{11}=0.
\label{E7}
\end{equation}
Thus even though the quantum 2D condition $G^{\mu\nu}(1)=0$ does not lead to any wave equation for the metric fluctuations,  the condition $\partial_{\nu}G^{\mu\nu}(2)=0$ does. And not only that, it precisely puts the fluctuations on the massless 2D light cone. The underlying conformal invariance of the theory thus leads to quantum graviton modes that are massless. In second order then the gravitational equations of motion of (\ref{E6}) exhibit a nice duality between the fermions and gravitons, with $G^{\mu\nu}$ and $T^{\mu\nu}_{\rm M}$ being both traceless and covariantly conserved, and with each containing modes that are massless. 

This duality of the fermion and graviton sectors has a consequence that is very significant for the zero-point problem. Specifically, since (\ref{E6}) is an identity, and since the quantized fermion sector has a non-vanishing zero-point energy density, it must be the case that the quantized graviton sector has a non-vanishing zero-point energy density too, and not only that, the graviton zero-point energy density must identically cancel that of the fermion. In fact this result is quite general. Specifically,  in any quantum gravity theory in any spacetime dimension in which the gravitational equations of motion are of the generic form given in (\ref{E6}), the graviton and matter field zero-point energy densities must cancel other. The cancellation will thus occur in any quantum gravity theory in which equations such as (\ref{E6}) remain meaningful in the presence of quantum corrections, i.e. in any theory of quantum gravity that is renormalizable. Since 4D Einstein gravity is not a renormalizable theory, in 4D Einstein gravity the cancellation would not immediately be expected to occur.

\subsection{Explicit Details of the 2D Cancellation Mechanism}
\label{S3}

With the flat space Dirac gamma matrices obeying $\gamma^{\mu}\gamma^{\nu}+\gamma^{\nu}\gamma^{\mu}=2\eta^{\mu\nu}$, and with $\eta^{00}$ being taken to be negative signatured in this paper,  in 2D we can set $\gamma^0=i\sigma_1$, $\gamma_0=-i\sigma_1$, $\gamma^1=\gamma_1=\sigma_2$, so that the free massless flat space 2D Dirac equation takes the form $(i\hbar\partial_0+i\hbar\sigma_3\partial_1)\psi(x,t)=0$. We define positive and negative energy solutions according to $u(k,\omega_k)e^{i(kx-\omega_kt)}$ and $v(-k,-\omega_k)e^{-i(kx-\omega_kt)}$ where for both solutions $\omega_k$ is defined to be the positive $|k|$. For these solutions normalized spinors are given by
\begin{eqnarray}
u(k,\omega_k)&=&\left(\begin{array}{c} 1 \\ 0 \end{array}\right),\qquad
u(-k,\omega_k)=\left(\begin{array}{c} 0 \\ 1 \end{array}\right),
\nonumber\\
v(-k,-\omega_k)&=&\left(\begin{array}{c} 1 \\ 0 \end{array}\right),\qquad
v(k,-\omega_k)=\left(\begin{array}{c} 0 \\ 1 \end{array}\right).
\label{E8}
\end{eqnarray}
For the fermion fields we introduce creation and annihilation operators according to
\begin{eqnarray}
\psi_{\alpha}^{\dagger}(x,t)&=&\int \frac{dk}{(2\pi)^{1/2}}\left[b^{\dagger}(k)u_{\alpha}^{\dagger}(k,\omega_k)e^{-i(kx-\omega_kt)}+d(k)v_{\alpha}^{\dagger}(-k,-\omega_k)e^{i(kx-\omega_kt)}\right],
\nonumber\\
\psi_{\alpha}(x,t)&=&\int \frac{dk}{(2\pi)^{1/2}}\left[b(k)u_{\alpha}(k,\omega_k)e^{i(kx-\omega_kt)}+d^{\dagger}(k)v_{\alpha}(-k,-\omega_k)e^{-i(kx-\omega_kt)}\right],
\label{E9}
\end{eqnarray}
Quantizing the fermion field according to $\{\psi_{\alpha}({ x},t),\psi_{\beta}^{\dagger}({ x}^{\prime},t)\}=\delta({ x}-{ x}^{\prime})\delta_{\alpha,\beta}$ then requires that its creation and annihilation operators obey
\begin{eqnarray}
&&\{b(k),b^{\dagger}(k^{\prime})\}=\delta(k-k^{\prime}),\qquad \{d(k),d^{\dagger}(k^{\prime})\}=\delta(k-k^{\prime}),\qquad \{b(k),b(k^{\prime})\}=0,
\nonumber \\
&&\{d(k),d(k^{\prime})\}=0,\qquad \{b^{\dagger}(k),b^{\dagger}(k^{\prime})\}=0,\qquad \{d^{\dagger}(k),d^{\dagger}(k^{\prime})\}=0,
\nonumber \\
&&\{b(k),d(k^{\prime})\}=0,\qquad \{b(k),d^{\dagger}(k^{\prime})\}=0,\qquad \{d(k),b^{\dagger}(k^{\prime})\}=0.
\label{E10}
\end{eqnarray}
Given the structure of (\ref{E9}), from (\ref{E5}) we can directly evaluate $T_{00}^{\rm M}=(i\hbar/2)\psi^{\dagger}\partial_0\psi-(i\hbar/2)[\partial_0{\psi}^{\dagger}]\psi$ to obtain
\begin{eqnarray}
T_{00}^{\rm M}&=&\frac{\hbar}{4\pi}\int dk \int  dk^{\prime}\bigg{[}(\omega_{k^{\prime}}-\omega_k)d(k)b(k^{\prime})v^{\dagger}(-k,-\omega_k)u(k^{\prime},\omega_{k^{\prime}})e^{i(k+k^{\prime})x}e^{-i(\omega_k+\omega_{k^{\prime}})t}
\nonumber \\
&-&(\omega_{k^{\prime}}-\omega_k)b^{\dagger}(k)d^{\dagger}(k^{\prime})u^{\dagger}(k,\omega_k)v(-k^{\prime},-\omega_{k^{\prime}})e^{-i(k+k^{\prime})x}e^{i(\omega_k+\omega_{k^{\prime}})t}
\nonumber \\
&+&(\omega_{k^{\prime}}+\omega_k)b^{\dagger}(k)b(k^{\prime})u^{\dagger}(k,\omega_k)u(k^{\prime},\omega_{k^{\prime}})e^{-i(k-k^{\prime})x}e^{i(\omega_k-\omega_{k^{\prime}})t}
\nonumber \\
&-&(\omega_{k^{\prime}}+\omega_k)d(k)d^{\dagger}(k^{\prime})v^{\dagger}(-k,-\omega_k)v(-k^{\prime},-\omega_{k^{\prime}})e^{i(k-k^{\prime})x}e^{-i(\omega_k-\omega_{k^{\prime}})t}\bigg{]}.
\label{E11}
\end{eqnarray}
Through the use of (\ref{E10}) we obtain a matter Hamiltonian $H_{\rm M}=\int_{-\infty}^{\infty} dx T^{\rm M}_{00}$ of the form
\begin{equation}
H_{\rm M }=\hbar\int_{-\infty}^{\infty} dk \omega_k\left[b^{\dagger}(k)b(k)-d(k)d^{\dagger}(k)\right],
\label{E12}
\end{equation}
and a zero-point energy density of the form
\begin{equation}
\langle \Omega | T^{\rm M}_{00}|\Omega \rangle =-\frac{\hbar}{2\pi}\int_{-\infty}^{\infty} dk \omega_k.
\label{E13}
\end{equation}
With  $T_{\mu\nu}^{\rm M}$ being traceless, the zero-point pressure evaluates to 
\begin{equation}
\langle \Omega | T^{\rm M}_{11}|\Omega \rangle =-\frac{\hbar}{2\pi}\int_{-\infty}^{\infty} dk \omega_k,
\label{E14}
\end{equation}
while the off-diagonal $\langle \Omega | T^{\rm M}_{01}|\Omega \rangle$ and the total vacuum momentum $P_1^{\rm M}=\int_{-\infty}^{\infty} dx \langle \Omega | T^{\rm M}_{01}|\Omega \rangle$ are both zero.

To quantize the gravitational field we need to find a quantization scheme for it that will produce a $G_{\mu\nu}(2)$ that possesses precisely the same structure as that found above for $T^{\rm M}_{\mu\nu}$, just as required by (\ref{E6}). Because the gravitational fluctuations have to obey (\ref{E7}), we see that given the trace condition $h=0$, at most only two of the three components of the 2D $h_{\mu\nu}$ can be independent degrees of freedom. Now the fact that  $h_{\mu\nu}$ would have two independent components is at first surprising since in a 2D theory one is free to make two general coordinate transformations, and thus one would expect to be able to reduce the initial 3-component $h_{\mu\nu}$ to just one independent component. However, coordinate transformations are precisely  that, i.e. they are precisely changes in classical coordinates of the classical form $x_{\mu} \rightarrow x_{\mu}+\epsilon_{\mu}$, , i.e. c-number not q-number transformations. While such transformations  could reduce the number of independent components of a classical $h_{\mu\nu}$, they cannot affect a quantum $h_{\mu\nu}$ since the coordinate transformation parameters $\epsilon_{\mu}$ are not quantum operators. Thus in the quantum theory all we can ask is whether we might  be able to reduce the number of quantum degrees of freedom by residual gauge transformations involving a quantum $\epsilon_{\mu}$ that preserve the solution given in (\ref{E7}). Thus if we set $\bar{h}_{\mu\nu}=h_{\mu\nu}+\partial_{\mu}\epsilon_{\nu}+\partial_{\nu}\epsilon_{\mu}$, $\bar{h}=h+2\partial_{\mu}\epsilon^{\mu}$ with a now quantum $\epsilon_{\mu}$, we need to require that $\epsilon_{\mu}$ obeys $\nabla^2\epsilon_{\mu}=0$, $\partial_{\mu}\epsilon^{\mu}=0$. These conditions can be satisfied by setting $\epsilon_{\mu}=f_{\mu}e^{i(kx-\omega_kt)}$ where $f_{\mu}$ has no spacetime dependence and obeys $k_{\mu}f^{\mu}=0$. With $f_{\mu}$ obeying this condition in 2D and with $k_{\mu}$ being lightlike, the only solution is that $f_{\mu}$ must be parallel to $k_{\mu}$ and be lightlike also. Under these conditions $\partial_{\mu}\epsilon_{\nu}+\partial_{\nu}\epsilon_{\mu}$ has to behave as $k_{\mu}k_{\nu}$, and thus has to be transverse. However, this is too restrictive a condition on the gauge transformation since an $h_{\mu\nu}$ that is not transverse for instance cannot be made transverse by it. There is thus no further gauge freedom. The fluctuation $h_{\mu\nu}$ thus has two independent components, which we take to be $h_{00}$ and $h_{01}$. 

As noted above,  gauge conditions that obey $\nabla^2\epsilon_{\mu}=0$, $\partial_{\mu}\epsilon^{\mu}=0$ leave $\partial_{\mu}h^{\mu\nu}$ invariant. With (\ref{E2}) explicitly showing that $G_{\mu\nu}(2)$ vanishes identically in transverse traceless modes that obey $\partial_{\mu}h^{\mu\nu}=0$, $h=0$, we thus see that in our 2D model, the gravitational fluctuations cannot be transverse traceless. Since there is a total of three independent conditions contained in $\partial_{\mu}h^{\mu\nu}=0$, $h=0$, and since the most general 2D $h_{\mu\nu}$ can only have three independent components, we see that by not having $h_{\mu\nu}$ be transverse traceless, we thus avoid the fact that there cannot be any transverse traceless gravitational fluctuations in 2D. While we refer to the degrees of freedom in $h_{\mu\nu}$ as graviton modes in this paper, by this we mean only that they are the quanta associated with the quantization of the gravitational field, and not that they are necessarily 2D analogs of the propagating transverse traceless modes of 4D Einstein gravity. In fact, with the 2D Einstein theory path integral being a constant, a 2D graviton cannot propagate at all, and thus has no need to be transverse.

With the solutions to (\ref{E7}) being massless plane waves with 2-vector $k^{\mu}=(\omega_k,k)$ and $\omega_k=|k|$ \cite{R5}, the most general form for $h_{00}$ and $h_{01}$ is given as 
\begin{eqnarray}
h_{00}(x,t)&=&\kappa_2\hbar^{1/2}\int \frac{dk}{(2\pi)^{1/2}(2\omega_k)^{1/2}}\left[A(k)e^{i(kx-\omega_kt)}+C(k)e^{-i(kx-\omega_kt)}\right]=h_{11}(x,t),
\nonumber\\
h_{01}(x,t)&=&\kappa_2\hbar^{1/2}\int \frac{dk}{(2\pi)^{1/2}(2\omega_k)^{1/2}}\left[B(k)e^{i(kx-\omega_kt)}+D(k)e^{-i(kx-\omega_kt)}\right],
\label{E15}
\end{eqnarray}
to give a total of four arbitrary field operators that are to be determined via (\ref{E6})   \cite{R6}.  In advance of actually doing a calculation one has no a priori sense as to what specific relations there might be between the various $A(k)$, $B(k)$, $C(k)$ and $D(k)$ or what commutation relations they might obey, since in the absence of there being any canonical quantization procedure for 2D Einstein gravity, the needed relations have to come from (\ref{E6}) alone, and thus have no need to look like any of the commutation relations that are familiarly encountered in the quantization of standard classical field theories. Thus for the moment we leave the four operators unspecified.

To determine $G_{00}(2)$, we insert (\ref{E15}) into (\ref{E2}) and obtain
\begin{eqnarray}
G_{00}(2)&=&\frac{\kappa_2^2\hbar}{16\pi}\int dk \int  dk^{\prime}\frac{(\omega_k k^{\prime}+k \omega_{k^{\prime}})}{(\omega_k\omega_{k^{\prime}})^{1/2}}
\nonumber \\
&&\times\bigg{(}[A(k),B(k^{\prime})]e^{i(k+k^{\prime})x}e^{-i(\omega_k+\omega_{k^{\prime}})t}
+[C(k),D(k^{\prime})]e^{-i(k+k^{\prime})x}e^{i(\omega_k+\omega_{k^{\prime}})t}
\nonumber \\
&&-[C(k),B(k^{\prime})]e^{-i(k-k^{\prime})x}e^{i(\omega_k-\omega_{k^{\prime}})t}
-[A(k),D(k^{\prime})]e^{i(k-k^{\prime})x}e^{-i(\omega_k-\omega_{k^{\prime}})t}\bigg{)}.
\label{E16}
\end{eqnarray}
As constructed, $G_{00}(2)$ has to obey $(1/\kappa_2^2)G_{00}(2)+T^{\rm M}_{00}=0$ where $T^{\rm M}_{00}$ is given in (\ref{E11}). Since $G_{00}(2)$ and $T^{\rm M}_{00}$ have both been expanded in complete bases of massless plane waves, simply by virtue of this completeness, there has to exist a solution to $(1/\kappa_2^2)G_{00}(2)+T^{\rm M}_{00}=0$  in which the gravitational field operators can be related to bilinear products of the fermion creation and annihilation operators. However, the information that we need for our purposes here can be extracted by taking vacuum matrix elements of this relation or by integrating it over all space. 

To extract some expectation value information, we take the two positive frequency operators $A(k)$ and $B(k)$ to annihilate the right vacuum $|\Omega \rangle$, and the two negative frequency operators $C(k)$ and $D(k)$ to annihilate the left vacuum $ \langle \Omega|$, and on requiring that the vacuum expectation values of the commutators $[C(k),B(k^{\prime})]$ and $[A(k),D(k^{\prime})]$ be given as
\begin{eqnarray}
&&\langle \Omega |[C(k),B(k^{\prime})]|\Omega \rangle=-\langle \Omega |B(k)C(k)|\Omega \rangle\delta(k-k^{\prime})=-f_{BC}(k)\delta(k-k^{\prime}),
\nonumber\\
&&\langle \Omega |[A(k),D(k^{\prime})]|\Omega \rangle=~~\langle \Omega |A(k)D(k)|\Omega \rangle\delta(k-k^{\prime})=~~f_{AD}(k)\delta(k-k^{\prime}),
\label{E17}
\end{eqnarray}
where $f_{BC}(k)$ and $f_{AD}(k)$ are c-number functions that are to be determined from the consistency of (\ref{E6}), we find that
\begin{eqnarray}
\frac{1}{\kappa_2^2}\langle \Omega | G^{00}(2)|\Omega \rangle &=& \frac{\hbar}{8 \pi }\int_{-\infty}^{\infty} dk k[f_{BC}(k)-f_{AD}(k)],
\nonumber\\
\frac{1}{\kappa_2^2}\langle \Omega | G^{01}(2)|\Omega \rangle &=& \frac{\hbar}{8 \pi }\int_{-\infty}^{\infty} dk \omega_k[f_{BC}(k)-f_{AD}(k)].
\label{E18}
\end{eqnarray}
To cancel the fermion  zero-point energy density, the gravitational field operators thus have to obey
\begin{eqnarray}
&&\frac{\hbar}{8 \pi }\int_{-\infty}^{\infty}dk k[f_{BC}(k)-f_{AD}(k)]
=\frac{\hbar}{2\pi}\int_{-\infty}^{\infty} dk \omega_k,
\nonumber\\
&&\frac{\hbar}{8 \pi }\int_{-\infty}^{\infty}dk \omega_k[f_{BC}(k)-f_{AD}(k)]
=0,
\label{E19}
\end{eqnarray}
a relation that, through our use of wave numbers rather than momenta, readily shows that both sides are of the same first order in $\hbar$. Finally, from (\ref{E19}), we see that the gravitational field must be quantized so that the quantities $f_{BC}(k)$ and $f_{AD}(k)$ obey just the one relation
\begin{equation}
k[f_{BC}(k)-f_{AD}(k)]=4\omega_k=4|k|.
\label{E20}
\end{equation}
In obtaining this condition, we thus achieve our primary purpose of showing how the zero-point energy densities of the gravitational and matter field mutually cancel each other identically, just as desired.

However, there is also information to be obtained from looking at the total energy rather than the energy density. Thus, on integrating $G_{00}(2)$ over all space and using (\ref{E6}) we obtain
\begin{eqnarray}
\frac{1}{\kappa_2^2}\int_{-\infty}^{\infty} dx G_{00}(2)&=&-\frac{\hbar}{4}\int_{-\infty}^{\infty}dk k\left([C(k),B(k)]+[A(k),D(k)]\right)
\nonumber\\
&=&-\hbar\int_{-\infty}^{\infty} dk \omega_k\left[b^{\dagger}(k)b(k)-d(k)d^{\dagger}(k)\right]
\label{E21}
\end{eqnarray}
a relation that holds for operators themselves rather than for vacuum matrix elements. If we now take $h_{00}$ and $h_{01}$ to be Hermitian just as one initially might expect of the gravitational field, and thus set $C(k)=A^{\dagger}(k)$, $D(k)=B^{\dagger}(k)$, we find that (\ref{E21}) then takes the form 
\begin{eqnarray}
&&-\frac{\hbar}{4}\int_{-\infty}^{\infty}dk k\left(A^{\dagger}(k)B(k)-B(k)A^{\dagger}(k)+A(k)B^{\dagger}(k)-B^{\dagger}(k)A(k)\right)
\nonumber\\
&&=-\hbar\int_{-\infty}^{\infty} dk \omega_k\left[b^{\dagger}(k)b(k)-d(k)d^{\dagger}(k)\right].
\label{E22}
\end{eqnarray}
However, rather than being Hermitian, the operator combination  $A^{\dagger}(k)B(k)-B(k)A^{\dagger}(k)+A(k)B^{\dagger}(k)-B^{\dagger}(k)A(k)$ is anti-Hermitian, and thus cannot satisfy (\ref{E22}). Quite remarkably then, we find that we cannot satisfy (\ref{E6}) with a Hermitian gravitational field. Since we shall meet precisely the same problem in the 4D conformal case, we defer discussion of this issue to Sec. \ref{S7}.

\subsection{The General Nature of the Zero-Point Problem}
\label{S4}

As constructed above, (\ref{E6}) actually addresses more than just the the zero-point energy problem, viz. that associated with the $(00)$ component of (\ref{E6}), as it actually addresses the entire zero-point fluctuation contribution, viz. that associated with all the other components of (\ref{E6}) as well. As noted in Sec. \ref{S3}, in the vacuum the fermion field contribution to the 2D $T^{\mu\nu}_{\rm M}$ is given by
\begin{equation}
\langle \Omega |T^{00}_{\rm M}|\Omega \rangle =\langle \Omega |T^{11}_{\rm M}|\Omega \rangle=-\frac{\hbar}{2\pi}\int_{-\infty}^{\infty} dk \omega_k,\qquad \langle \Omega |T^{01}_{\rm M}|\Omega \rangle =0,
\label{E23}
\end{equation}
as summed over negative energy fermion modes with 2-vector $k^{\mu}=(\omega_k,k)$ and $\omega_k=|k|$. As  such, we see that $\langle \Omega |T^{\mu\nu}_{\rm M}|\Omega \rangle$ behaves as a traceless 2D perfect fluid $(\rho_{\rm M}+p_{\rm M})U^{\mu}U^{\nu}+p_{\rm M}\eta^{\mu\nu}$ with a zero-point pressure $p_{\rm M}$ and a zero-point energy density $\rho_{\rm M}$ that are equal to each other and given by the quadratically divergent $p_{\rm M}=\rho_{\rm M}= -(\hbar/2\pi)\int_{-\infty}^{\infty} dk \omega_k$. Quantization of the fermion field thus gives both a zero-point energy density and a zero-point pressure. Moreover, this form for $T^{\mu\nu}_{\rm M}$ cannot be associated with a cosmological constant term, as a cosmological constant is associated with a perfect fluid $T^{\mu\nu}_{\Lambda}=-\Lambda \eta^{\mu\nu}$ whose pressure has the opposite sign to that of its energy density and whose trace is the non-vanishing $\eta_{\mu\nu}T^{\mu\nu}_{\Lambda}=-2\Lambda$ (in 2D). With the zero-point pressure in $T^{\mu\nu}_{\rm M}$ having the same sign as the zero-point energy density, and with $T^{\mu\nu}_{\rm M}$ being traceless, we see that the zero-point and cosmological constant problems are intrinsically different, with there actually being two vacuum problems that need to be solved not just one, namely both the cosmological constant and the zero-point problems as they are in principle different. 

As we show below in Sec. \ref{S5}, when we give a flat spacetime fermion a mass by a mean-field type symmetry breaking, we will induce a cosmological constant term and generate changes in  the above zero-point $T^{\mu\nu}_M$. With the fermion mass-energy relation changing to $k^{\mu}=((k^2+m^2/\hbar^2)^{1/2},k)$, the zero-point $\langle \Omega |T^{\mu\nu}_{\rm M}|\Omega \rangle$ will acquire two mass-dependent contributions, one being logarithmically divergent and the other being finite, while the induced cosmological constant will be a logarithmically divergent term that has no finite part. Since the overall trace of the total energy-momentum tensor will continue to vanish in a conformal invariant theory even after symmetry breaking (changing the vacuum does not affect Ward identities), the zero-point logarithmic divergence will be cancelled by the induced cosmological constant term. The mass-independent quadratic divergence and the mass-dependent finite part of the massive $\langle \Omega |T^{\mu\nu}_{\rm M}|\Omega \rangle$ will need to be cancelled by the graviton zero-point term (in a way that we describe below). By recognizing the distinction between the zero-point contribution and the cosmological constant contribution, we are able to monitor both contributions and show how all these various contributions coordinate to mutually cancel each other, doing so no matter how big these contributions might be.

Having drawn the above distinction between the zero-point energy density term and the cosmological constant term, we need to reconcile our analysis with the well-known theorem that the energy of a Lorentz-invariant vacuum must be zero (a theorem that is to hold in any spacetime dimension). Since the zero-point energy was calculated above to be infinite, and thus seemingly in violation of the theorem, it is thought that the theorem is restored via the renormalization of the vacuum energy density, a procedure that typically leads to renormalization anomalies such as the trace anomaly. We have used the expression "seemingly in violation" here since the calculation of the zero-point energy given above is a strictly Lorentz invariant one in which Lorentz invariance is maintained at every step and yet the resulting vacuum energy $E$ that obeys $E|\Omega\rangle= \int dx T_{00}|\Omega\rangle$ is not found to be zero. Moreover, one can even write the vacuum energy density difference between massive and massless free fermions in the manifestly covariant form $\epsilon(m)-\epsilon(m=0)=(i/\hbar)\int d^Dp/(2\pi)^D[{\rm Tr Ln}(\gamma^{\mu}p_{\mu}-m+i\epsilon)-{\rm Tr Ln}(\gamma^{\mu}p_{\mu}+i\epsilon)]$ (in general dimension D), to yield a manifestly covariant quantity that explicitly evaluates to minus infinity and not to zero. (In fact is the very non-vanishing of energy density differences such as these that serves as the basis for establishing that dynamical symmetry breaking in Nambu-Jona-Lasinio type models actually occurs.)

To resolve the issue we need to recall the derivation of the zero vacuum energy theorem. Specifically, since the momentum of the vacuum $P_i|\Omega\rangle=\int dx T_{0i}|\Omega \rangle$ is strictly zero,  under a Lorentz boost with $\gamma=(1-v^2/c^2)^{-1/2}$ the energy of a Lorentz invariant vacuum would transform into $\gamma E$ and thus have to obey $E=\gamma E$. Now ordinarily one would expect a relation such as this to admit of only one solution, namely that with $E=0$, since one assumes that the vacuum energy is finite. However, there is another way to satisfy the $E=\gamma E$ relation, namely by having $E$ be infinite. In the calculation of the vacuum energy density given above the vacuum that was used is not empty (for a truly empty vacuum $E$ would indeed be zero). Rather it is filled to the brim with all the available negative energy fermion modes, i.e. filled with an infinite number of such modes. Since the set of modes is infinite in number, under a Lorentz boost a state of any given momentum will transform into some other state in the set since no momenta values are absent. The configuration consisting of all of the modes taken together is thus boost invariant, to thereby provide a many-body Lorentz invariant vacuum configuration that is not required to have zero energy. The requirement that a Lorentz invariant vacuum have zero energy thus does not apply to vacuum configurations with an infinite number of degrees of freedom.

Now suppose we have such an infinite zero-point energy and try to renormalize it so as to make it finite. To this end we would first regulate the momentum integral in (\ref{E23}) by a procedure such as introducing a cut-off, with  the allowed momenta values then being cut off at some large value $K$. In so doing we would immediately violate both Lorentz invariance and scale invariance, Lorentz invariance since now boosts that would boost states to momenta greater than $K$ would not leave the cut-off configuration of modes invariant, and scale invariance since the very introduction of $K$ introduces an intrinsic scale. While the Lorentz invariance is restored after the renormalization procedure is carried through, the scale invariance is not and trace anomalies result. However, the emergence of a trace anomaly is due only to the fact that we sought to regulate the vacuum energy density in the first place. No such regularization concerns would be encountered at all if we could cancel the matter field energy density by  that of some other field. However, such a cancellation would have to be a complete one since the vacuum energy of a Lorentz invariant vacuum could not be finite, with $E=0$ being the only alternative to $E=\infty$. (Even if the cancellation through other fields was not complete, the vacuum energy could still be brought to zero by a cosmological constant, but if we are to restrict to cancellations by fields alone, they would have to bring about a total cancellation.) While there are various ways to achieve a total cancellation (such as supersymmetry), as we have seen above, in renormalizable theories of gravity the zero-point energy density of the gravitational field can also serve this purpose. And indeed, since the energy-momentum tensor of the matter field is defined as $T_{\rm M}^{\mu\nu}=2(\delta I_{\rm M}/\delta g_{\mu\nu})/(-g)^{1/2}$, we can analogously define a gravitational energy-momentum tensor as $T_{\rm GRAV}^{\mu\nu}=2(\delta I_{\rm GRAV}/\delta g_{\mu\nu})/(-g)^{1/2}$ (as symmetrized as necessary) and reinterpret (\ref{E6}) as the generic
\begin{equation}
T_{\rm UNIV}^{\mu\nu}=T_{\rm GRAV}^{\mu\nu}+T_{\rm M}^{\mu\nu}=0.
\label{E24}
\end{equation}
As long as equations such as (\ref{E24}) remain meaningful after radiative corrections (i.e. as long as they are not destroyed by  counterterms that are not controllable), the import of (\ref{E24}) is that the total energy-momentum tensor of the universe is zero and that gravitational zero-point and the matter field zero-point energy densities must cancel each other identically. Thus if the theory associated with $I_{\rm GRAV}+I_{\rm M}$ is renormalizable, the $T_{\rm UNIV}^{\mu\nu}=0$ condition is not modified by either gravitational field or matter field radiative corrections, to thus ensure that all infinities in $T_{\rm GRAV}^{\mu\nu}$ and $T_{\rm M}^{\mu\nu}$ must mutually cancel each other identically. And not only must all infinities cancel, all finite parts must cancel as well, to thus keep the total energy-momentum tensor of the universe  $T_{\rm UNIV}^{\mu\nu}$ zero. Since the vanishing of $T_{\rm UNIV}^{\mu\nu}$ follows only from stationarity with respect to the metric, the condition that $T_{\rm UNIV}^{\mu\nu}$ be zero is the covariant generalization of the condition that Lorentz invariant vacua have zero energy. The solution to the matter field zero-point energy problem then is to include the zero-point energy of the gravitational field as well. Moreover, with such an inclusion, not only is the zero-point fluctuation problem solved, no regularization of the matter field zero-point energy is needed, and the consistency of the zero-point sector does not give rise to a trace anomaly. 

\subsection{Mass Generation and the Cosmological Constant Problem}
\label{S5}

To generate a fermion mass in our 2D model we introduce a four-fermi coupling analogous to the 4D one of Nambu and Jona-Lasinio, to give a total universe action of the form
\begin{equation} 
I_{\rm UNIV}=I_{\rm GRAV}+I_{\rm M}+\frac{g}{2}\int d^2x(-g)^{1/2}[\bar{\psi}\psi]^2,
\label{E25}
\end{equation}
where $I_{\rm GRAV}$ and $I_{\rm M}$ are as given in (\ref{E4}). With the four-fermi coupling constant $g$ being dimensionless in 2D, we find that under a local conformal transformation in which the metric, the vierbeins and the fermion respectively transform as $g_{\mu\nu}(x)\rightarrow e^{2\alpha(x)}g_{\mu\nu}(x)$, $V_{\mu}^a(x)\rightarrow e^{\alpha(x)}V_{\mu}^a(x)$, $\psi(x)\rightarrow e^{-\alpha(x)/2}\psi(x)$, the action $I_{\rm UNIV}$ is left invariant. The action of (\ref{E25}) is thus just the conformal invariant action we need. In a flat spacetime background the fermion equation of motion is given by  $i\hbar\gamma^{\mu}\partial_{\mu}\psi -g[\bar{\psi}\psi]\psi=0$, and in solutions to this equation of motion the fermion energy-momentum tensor is given by the traceless (in 2D)
\begin{equation} 
T^{\mu\nu}_{\rm M}=i\hbar\bar{\psi}\gamma^{\mu}\partial^{\nu}\psi -\eta^{\mu\nu}(g/2)[\bar{\psi}\psi]^2,\qquad 
\eta_{\mu\nu}T^{\mu\nu}_{\rm M}=i\hbar\bar{\psi}\gamma^{\mu}\partial_{\mu}\psi -g[\bar{\psi}\psi]^2=0.
\label{E26}
\end{equation}
(In (\ref{E26}) the term $i\hbar\bar{\psi}\gamma^{\mu}\partial^{\nu}\psi$ is a shorthand for the $T^{\mu\nu}_{\rm M}$ term given in (\ref{E5}).) To now explore dynamical symmetry breaking in theory, we shall follow Nambu and Jona-Lasinio and use the mean-field approximation.

In the mean-field, Hartree-Fock approximation one looks for self-consistent, translation invariant states $|S\rangle$ in which $\langle S|\bar{\psi}\psi | S\rangle= \langle S|\psi^{\dagger}i\sigma_1\psi | S\rangle=im/g$ and  $\langle S|(\bar{\psi}\psi -im/g)^2| S\rangle =0$ (for our choice of spacetime metric $\gamma^0=i\sigma_1$ is anti-Hermitian), with the parameter $m$ being independent of the spacetime coordinates. In such states the fermion equation of motion and the mean-field fermion energy-momentum tensor $T^{\mu\nu}_{\rm MF}$  take the form 
\begin{eqnarray}
&&i\hbar\gamma^{\mu}\partial_{\mu}\psi -im\psi=0,\qquad \langle S|T^{\mu\nu}_{\rm MF}| S\rangle=\langle S|i\hbar\bar{\psi}\gamma^{\mu}\partial^{\nu}\psi| S\rangle +\frac{m^2}{2g}\eta^{\mu\nu}, 
\nonumber\\
&&\eta_{\mu\nu}\langle S|T^{\mu\nu}_{\rm MF}| S\rangle=im\langle S|\bar{\psi}\psi| S\rangle +\frac{m^2}{g}=0,
\label{E27}
\end{eqnarray}
with the mean-field approximation preserving tracelessness. In conformal invariant theories then, we see that, just as noted in \cite{R4}, one can have mass generation without the trace needing to be non-zero. (It is only mechanical masses that are associated with a non-traceless energy-momentum tensor, but not dynamical ones.) With the emergence of the $(m^2/2g)\eta^{\mu\nu}$ term in (\ref{E27}), we thus see that dynamical mass generation induces a mean-field cosmological constant term with $\Lambda_{\rm MF} =-m^2/2g$.

On evaluating the quantity $\langle S|i\hbar\bar{\psi}\gamma^{\mu}\partial^{\nu}\psi| S\rangle$ in solutions to the now massive Dirac equation, we find it to be given by
\begin{eqnarray}
&& \langle S|i\hbar\bar{\psi}\gamma^{0}\partial^{0}\psi| S\rangle =-\frac{\hbar}{2\pi}\int_{-\infty}^{\infty} dk \omega_k,\qquad \langle S|i\hbar\bar{\psi}\gamma^{0}\partial^{1}\psi| S\rangle =0,\
 \nonumber\\
&& \langle S|i\hbar\bar{\psi}\gamma^{1}\partial^{1}\psi| S\rangle =-\frac{\hbar}{2\pi}\int_{-\infty}^{\infty} dk \frac{k^2}{\omega_k},
\label{E28}
\end{eqnarray}
i.e. just as in (\ref{E23}) except that now $\omega_k$ is given by the massive $\omega_k=(k^2+m^2/\hbar^2)^{1/2}$. 
Similarly, on introducing a convenient momentum cut-off $K$ to parameterize the degree of divergence, from the trace condition one obtains 
\begin{equation}
im\langle S|\bar{\psi}\psi| S\rangle =\frac{\hbar}{2\pi}\int_{-K}^{K} dk\left( \frac{m^2}{\hbar^2\omega_k}\right)=\frac{m^2}{\pi\hbar}{\rm ln}\left(\frac{2\hbar K}{m}\right)=
-\frac{m^2}{g}=2\Lambda_{\rm MF} ,
\label{E29}
\end{equation}
the self-consistent gap equation for the fermion mass, with familiar solution
\begin{equation}
m=2\hbar Ke^{\pi\hbar/g}.
\label{E30}
\end{equation}

Comparing now with the massless fermion case discussed earlier (equivalent in the present context to a Hartree-Fock approximation to (\ref{E26}) in a normal vacuum $|N\rangle$ in which $\langle N|\bar{\psi}\psi | N\rangle =0$), we see that in the massless case the vanishing trace condition is satisfied by having $\eta_{\mu\nu} \langle N|i\hbar\bar{\psi}\gamma^{\mu}\partial^{\nu}\psi| N\rangle$ and  $\langle N|\bar{\psi}\psi | N\rangle $ both be zero. Then when we change to the vacuum $| S\rangle $, $ \eta_{\mu\nu} \langle S|i\hbar\bar{\psi}\gamma^{\mu}\partial^{\nu}\psi| S\rangle$ and $\langle S|\bar{\psi}\psi | S\rangle $ both become logarithmic divergent, doing so in a way that, as previously noted in  \cite{R7},  enables the two quantities to cancel each other identically in the vanishing trace. Thus in the normal vacuum  $|N\rangle$ the vanishing trace condition is satisfied trivially, but in the self-consistent vacuum  $|S\rangle$ the vanishing trace condition is satisfied non-trivially. That such a cancellation can take place at all is due to the fact that both of the $ \eta_{\mu\nu} \langle S|i\hbar\bar{\psi}\gamma^{\mu}\partial^{\nu}\psi| S\rangle$ and $\langle S|\bar{\psi}\psi | S\rangle$ matrix elements are evaluated in one and the same state $| S\rangle $, with the two matrix elements thus knowing about each other. This is to be contrasted with theories in which there is a fundamental cosmological constant term, since the constant value that a fundamental $\Lambda$ takes is fixed once and for all and does not adjust to whatever states might be occupied. The ability of the four-fermi theory to control the cosmological constant is thus due to the fact that $\Lambda_{\rm MF} $  is dynamically determined and adjusts to whatever states are occupied. Working with different states (such as the time- or spatially-dependent coherent states we consider below in Sec. \ref{S9}) will require yet further adjustment, but will still leave the vanishing trace condition and its associated cancellations untouched. In theories in which $\Lambda$ is fundamental, such a $\Lambda$ is not affected by which matter field states might be occupied in the matter field energy-momentum tensor, to thus make it very difficult to solve the cosmological constant problem in theories with a fundamental $\Lambda$. However when one associates the cosmological constant with a dynamically induced, state-dependent $\Lambda_{\rm MF}$ term, one then does have control over it. This then is the power of dynamical symmetry breaking.

In achieving the above cancellation in the matter field trace, gravity played no role. Specifically, no contribution from the gravitational field is needed since the vanishing of the matter field trace is secured by the matter field equations of motion alone without reference to those obeyed by the gravitational field at all. When the gravitational field sector $G_{\mu\nu}(2)$ being independently traceless (c.f. (\ref{E2})), the gravitational field equations of (\ref{E6}), as now written in the form
\begin{equation}
\frac{1}{\kappa_2^2}G^{\mu\nu}+T^{\mu\nu}_{\rm MF}=0,
\label{E31}
\end{equation}
can be consistently imposed without at all affecting either the vanishing matter field trace or the implications of its vanishing. Symmetry breaking in the fermion sector does not affect the graviton mass, to thus leave the graviton massless and  the $\eta^{\mu\nu}G_{\mu\nu}(2)=0$ condition untouched. 

While we have been able to cancel the logarithmic divergences in $ \eta_{\mu\nu} \langle S|i\hbar\bar{\psi}\gamma^{\mu}\partial^{\nu}\psi| S\rangle$ and $\langle S|\bar{\psi}\psi | S\rangle $ without reference to gravity at all, there is a remaining quadratic divergence in the untraced $ \langle S|i\hbar\bar{\psi}\gamma^{\mu}\partial^{\nu}\psi| S\rangle$. To characterize and deal with this divergence we note that as constructed, the mean-field $\langle S|i\hbar\bar{\psi}\gamma^{\mu}\partial^{\nu}\psi| S\rangle$ has the form of a matter field perfect fluid $(\rho_{\rm MF}+p_{\rm MF})U^{\mu}U^{\nu}+p_{\rm MF}\eta^{\mu\nu}$ where $\rho_{\rm MF}$ and $p_{\rm MF}$ are given by
\begin{eqnarray}
\rho_{\rm MF} &=&-\frac{\hbar}{2\pi}\int_{-K}^{K}  dk\omega_k=-\frac{\hbar}{2\pi}\left[K^2+\frac{m^2}{2\hbar^2}+\frac{m^2}{\hbar^2}{\rm ln}\left(\frac{2\hbar K}{m}\right)\right],
\nonumber\\
p_{\rm MF}&=&-\frac{\hbar}{2\pi}\int_{-K}^{K}  dk\frac{k^2}{\omega_k}=-\frac{\hbar}{2\pi}\left[K^2+\frac{m^2}{2\hbar^2}-\frac{m^2}{\hbar^2}{\rm ln}\left(\frac{2\hbar K}{m}\right)\right].
\label{E32}
\end{eqnarray}
Thus with the induced cosmological constant term, the full $\langle S|T^{\mu\nu}_{\rm MF}| S\rangle$ can be written as
\begin{eqnarray}
\langle S|T^{\mu\nu}_{\rm MF}| S\rangle&=&(\rho_{\rm MF}+p_{\rm MF})U^{\mu}U^{\nu}+p_{\rm MF}\eta^{\mu\nu} -\eta^{\mu\nu}\Lambda_{\rm MF},
\nonumber\\
\eta_{\mu\nu} \langle S|T^{\mu\nu}_{\rm MF}| S\rangle&=&p_{\rm MF}-\rho_{\rm MF} -2\Lambda_{\rm MF}=0,
\label{E33}
\end{eqnarray}
where $\Lambda_{\rm MF} $ is given as in (\ref{E29}), i.e. as
\begin{equation}
\Lambda_{\rm MF} =\frac{m^2}{2\pi\hbar}{\rm ln}\left(\frac{2\hbar K}{m}\right).
\label{E34}
\end{equation}
Through use of the vanishing trace condition it is very convenient to  eliminate $\Lambda_{\rm MF} $, since $\langle S|T^{\mu\nu}_{\rm MF}| S\rangle$ can then be written in the manifestly traceless (in 2D) form
\begin{equation}
\langle S|T^{\mu\nu}_{\rm MF}| S\rangle=(\rho_{\rm MF}+p_{\rm MF})\left[U^{\mu}U^{\nu}+\frac{1}{2}\eta^{\mu\nu}\right],
\label{E35}
\end{equation}
where
\begin{equation}
\rho_{\rm MF}+p_{\rm MF}
=-\frac{\hbar}{2\pi}\int_{-K}^{K}  dk\left[2\omega_k-\frac{m^2}{\hbar^2\omega_k}\right]
=-\frac{\hbar}{\pi}\left(K^2+\frac{m^2}{2\hbar^2}\right),
\label{E36}
\end{equation}
with the mass-dependent logarithmic divergences in $\rho_{\rm MF}$, $p_{\rm MF}$ and $\Lambda_{\rm MF}$ all having cancelled each other identically.

Since $G_{\mu\nu}(2)$ is traceless, in the vacuum $|\Omega \rangle$ (here $|\Omega \rangle$ is normal with respect to graviton field bilinears and only self-consistent with respect to fermion bilinears),  its matrix element $\langle \Omega | G_{\mu\nu}(2)|\Omega \rangle/\kappa_2^2$ must have the same generic traceless form as $\langle S|T^{\mu\nu}_{\rm MF}| S\rangle$ and be given by 
\begin{equation}
\frac{1}{\kappa_2^2}\langle \Omega|G^{\mu\nu}(2)|\Omega \rangle=(\rho_{\rm GRAV}+p_{\rm GRAV})\left[U^{\mu}U^{\nu}+\frac{1}{2}\eta^{\mu\nu}\right],\qquad p_{\rm GRAV}-\rho_{\rm GRAV}=0.
\label{E37}
\end{equation}
Comparing with (\ref{E18}) we see that to cancel the quadratic divergence in (\ref{E36}) the quantities $f_{BC}(k)$ and $f_{AD}(k)$ as defined in (\ref{E17})  via the gravitational field commutators have to obey 
\begin{equation}
\frac{\hbar}{8 \pi }\int_{-K}^{K} dk k[f_{BC}(k)-f_{AD}(k)]
=\frac{\hbar}{2\pi}\int_{-K}^{K}  dk\left[\omega_k-\frac{m^2}{2\hbar^2\omega_k}\right]
=\frac{\hbar}{2\pi}\left(K^2+\frac{m^2}{2\hbar^2}\right).
\label{E38}
\end{equation}

Now at first sight it would appear that (\ref{E38}) could not possibly be satisfied. Specifically, from the consistency of the massless fermion case discussed above, via (\ref{E19}) we had already determined that the quantity that appears on the left-hand side  of (\ref{E38}) had to obey $k[f_{BC}(k)-f_{AD}(k)]=4|k|$. And with the graviton remaining massless, there would appear to be no way to cancel the mass-dependent $m^2/4\pi \hbar$ term present in (\ref{E38}). While the massless graviton can thus readily continue to cancel the mass-independent quadratic divergence in $\langle S|T^{\mu\nu}_{\rm MF}| S\rangle$, it would appear that it could not cancel the finite part. Despite this however, as we saw in our discussion of the implications of  the generic (\ref{E24}), once all the divergent parts have been cancelled, no non-vanishing finite part could be left over since the total $T^{\mu\nu}_{\rm UNIV}$ of the universe has to vanish identically. The cancellation required by (\ref{E38}) thus has to take place.

To explain how it does, we note that there is a crucial difference in how (\ref{E6}) and (\ref{E31}) are to be satisfied. In (\ref{E6}) the gravitational and fermion matter fields are both massless, and (\ref{E6}) relates two complete massless mode bases, to lead to (\ref{E39}). However, in (\ref{E31}) the graviton is massless while the fermion is massive, with (\ref{E31}) relating two different plane wave mode bases, one massless, the other massive. Now since both bases are complete, they can still be related. To this end we recall the comparison of massless and massive free fermion mode bases given in \cite{R3}, where the bases are related by a Bogoliubov transform, with a massive fermion mode being given as a linear combination of particle and hole modes  of the massless fermion. In such a transform the transformation coefficients depend on the fermion mass parameter $m$ even while the massless mode wave functions themselves do not. A similar situation thus obtains in the gravitational case, with (\ref{E31}) requiring that the 2D gravitational field operators $A(k)$, $B(k)$, $C(k)$ and $D(k)$ as defined in (\ref{E15}) have to depend on the mass parameter $m$. As long as the modes themselves are massless plane waves, the gravitational fluctuations will obey the massless wave equations given in (\ref{E7})  no matter what form the operators  $A(k)$, $B(k)$, $C(k)$ and $D(k)$ might specifically take. All that is needed of them is that their commutators are given as in (\ref{E17}), except that now the coefficients $f_{BC}(k)$ and $f_{AD}(k)$ have to obey (\ref{E38}), and must thus now obey
\begin{equation}
k[f_{BC}(k)-f_{AD}(k)]=4\left[(k^2+m^2)^{1/2}-\frac{m^2}{2\hbar^2(k^2+m^2)^{1/2}}\right].
\label{E39}
\end{equation}
With this requirement on the quantization of a gravitational field coupled to a massive fermion, the total $T^{\mu\nu}_{\rm UNIV}$ of the universe does indeed then vanish identically, just as required. 

Since the 2D Einstein gravitational field cannot be quantized canonically, the gravitational field has to instead be quantized by its coupling to a quantized fermion, with different gravitational field quantization conditions emerging dependent on whether the fermion is massless or massive. To keep $T^{\mu\nu}_{\rm UNIV}$ zero then, the coupling to the fermion thus quanitizes the gravitational field according to whatever conditions are required, doing so in a way in which the zero-point and cosmological constant problems are then mutually resolved. While our 2D quantum Einstein gravity model is only a toy model, it does capture the essence of how conformal invariance and stationarity with respect to the metric (c.f.  (\ref{E24})) can control and resolve the zero-point and cosmological constant problems. Armed with this information, we proceed now to discuss a more realistic gravitational theory with a natural tracelessness constraint, one in which there now is gravitational scattering, namely conformal gravity in 4D.

\section{The Four-Dimensional Case}
\label{U2}

\subsection{4D Conformal Gravity}
\label{S6}

In four spacetime dimensions the requirement that a general coordinate invariant gravitational  action also be invariant under local conformal transformations of the form $g_{\mu\nu}(x) \rightarrow e^{2\alpha(x)}g_{\mu\nu}(x)$ leads to a unique gravitational action, the conformal Weyl action given by
\begin{eqnarray}
I_{\rm W}&=&-\alpha_g\int d^4x (-g)^{1/2}C_{\lambda\mu\nu\kappa}
C^{\lambda\mu\nu\kappa}
\nonumber \\
&=&-\alpha_g\int d^4x (-g)^{1/2}\left[R_{\lambda\mu\nu\kappa}
R^{\lambda\mu\nu\kappa}-2R_{\mu\kappa}R^{\mu\kappa}+\frac{1}{3}
(R^{\alpha}_{\phantom{\alpha}\alpha})^2\right].
\label{E40}
\end{eqnarray}
Here $\alpha_g$ is a dimensionless gravitational coupling constant, and $C_{\lambda\mu\nu\kappa}$ is the Weyl or conformal tensor \cite{R8}
\begin{equation}
C_{\lambda\mu\nu\kappa}= R_{\lambda\mu\nu\kappa}
-\frac{1}{2}\left(g_{\lambda\nu}R_{\mu\kappa}-
g_{\lambda\kappa}R_{\mu\nu}-
g_{\mu\nu}R_{\lambda\kappa}+
g_{\mu\kappa}R_{\lambda\nu}\right)
+\frac{1}{6}R^{\alpha}_{\phantom{\alpha}\alpha}\left(
g_{\lambda\nu}g_{\mu\kappa}-
g_{\lambda\kappa}g_{\mu\nu}\right).
\label{E41}
\end{equation}
The Weyl tensor has a remarkable geometric property, namely that under a local conformal transformation the tensor $C^{\lambda}_{\phantom{\lambda}\mu\nu\kappa}$ transforms into itself. Thus even though the Riemann tensor and the Ricci tensor and scalar acquire second derivatives of  $\alpha(x)$ under a local conformal transformation,  in the particular linear combination given in (\ref{E41}), all derivatives of $\alpha(x)$ identically drop out.  Since all gauge function derivatives also drop out of the Maxwell tensor $F_{\mu\nu}$ when it is transformed by a gauge transformation, the Weyl tensor bears the same relation to conformal transformations as the Maxwell tensor does to gauge transformations. For an arbitrary local complex transformation on a generic matter field of the form $\psi(x)\rightarrow e^{\alpha(x)+i\beta(x)}\psi(x)$, electromagnetism gauges the imaginary part of the phase and conformal gravity gauges the real part. As constructed, (\ref{E40}) is the conformal analog of the Maxwell action $-(1/4)\int d^4 x(-g)^{1/2}F_{\mu\nu}F^{\mu\nu}$, and the kinematic relation $g^{\mu\kappa}C^{\lambda}_{\phantom{\lambda}\mu\nu\kappa}=0$ that the Weyl tensor obeys (the Weyl tensor being the traceless piece of the Riemann tensor) is the counterpart of the kinematic relation  $g^{\mu\nu}F_{\mu\nu}=0$ obeyed by the Maxwell tensor.  Conformal gravity thus endows gravity with a structure very similar to that found in gauge theories, and since the energy-momentum tensor of these very same $SU(3)\times SU(2) \times U(1)$ gauge theories is to serve as the source of gravity, it is both natural and unifying to give gravity an analogous such structure.  

For our purposes here, the virtues of using $I_{\rm W}$ as the 4D gravitational action rather than the standard Einstein-Hilbert action $I_{\rm EH}=-(1/2\kappa_4^2)\int d^4x (-g)^{1/2}R^{\alpha}_{\phantom{\alpha}\alpha}$ are threefold. Firstly, unlike the standard Einstein theory, in the conformal theory one is not permitted to add any fundamental cosmological constant term to the action as it would violate the underlying conformal invariance of the conformal theory. Conformal gravity thus has a control over the cosmological constant that standard theory does not, and thus provides a good starting point for attacking the cosmological constant problem. Secondly, variation of $I_{\rm W}$ with respect to the metric, viz.
\begin{equation}
\frac{1}{(-g)^{1/2}}\frac{\delta I_{\rm W}}{ \delta g_{\mu
\nu}}=-2\alpha_gW^{\mu \nu}
\label{E42}
\end{equation}
yields a rank two gravitational tensor $W^{\mu \nu}$ that is kinematically traceless, to thus enable us to recover the key features of the 2D analysis that was given above. Thirdly, since $\alpha_g$ is dimensionless, as a quantum theory conformal gravity is power counting renormalizable, and thus unlike standard Einstein gravity, conformal gravity is quantum-mechanically sensible. Now since conformal gravity is based on fourth-order rather than second-order derivative functions of the metric, there has been a concern that the theory might not be unitary since higher-derivative theories have been thought to possess negative Dirac norm ghost states. However these concerns have now been resolved \cite{R1,R2}, with it having been shown in \cite{R1,R2} that in higher-derivative theories one should not use the standard Dirac norm. Rather, one should use the $\cal{C}\cal{P}\cal{T}$ norm of $\cal{P}\cal{T}$ theories \cite{R9}, and that when one does, there are then no ghost states at all and the theory is unitary. We shall discuss the work of \cite{R1,R2} in more detail below in Secs. \ref{S7} and \ref{S8}, noting for the moment only that since conformal gravity is a consistent theory of quantum gravity in 4D, one can thus viably explore both the cosmological constant and zero-point problems in it.

When conformal gravity is treated as a classical theory, it is found \cite{R10} that the Lanzcos Lagrangian
\begin{equation}
L_{\rm L}=(-g)^{1/2}\left[R_{\lambda\mu\nu\kappa}
R^{\lambda\mu\nu\kappa}-4R_{\mu\kappa}R^{\mu\kappa}+
(R^{\alpha}_{\phantom{\alpha}\alpha})^2\right]
\label{E43}
\end{equation}
is a total derivative. At the classical level one can thus simplify the Weyl action to the form
\begin{equation}
I_{\rm W}=-2\alpha_g\int d^4x
(-g)^{1/2}\left[R_{\mu\kappa}R^{\mu\kappa}-\frac{1}{3}
(R^{\alpha}_{\phantom{\alpha}\alpha})^2\right],
\label{E44}
\end{equation}
with its variation in the presence of a matter action $I_{\rm M}$ then leading (see e.g. \cite{R4} and references therein for details) to the equation of motion
\begin{equation}
4\alpha_g W^{\mu\nu}=4\alpha_g\left[W^{\mu
\nu}_{(2)}-\frac{1}{3}W^{\mu\nu}_{(1)}\right]=T^{\mu\nu}_{\rm M},
\label{E45}
\end{equation}
where $W^{\mu \nu}_{(1)}$ and $W^{\mu \nu}_{(2)}$ are respectively given by
\begin{equation}
W^{\mu \nu}_{(1)}= 
2g^{\mu\nu}(R^{\alpha}_{\phantom{\alpha}\alpha})          
^{;\beta}_{\phantom{;\beta};\beta}                                              
-2(R^{\alpha}_{\phantom{\alpha}\alpha})^{;\mu;\nu}                           
-2 R^{\alpha}_{\phantom{\alpha}\alpha}
R^{\mu\nu}                              
+\frac{1}{2}g^{\mu\nu}(R^{\alpha}_{\phantom{\alpha}\alpha})^2,
\label{E46}
\end{equation}                                 
and 
\begin{equation}
W^{\mu \nu}_{(2)}=
\frac{1}{2}g^{\mu\nu}(R^{\alpha}_{\phantom{\alpha}\alpha})   
^{;\beta}_{\phantom{;\beta};\beta}+
R^{\mu\nu;\beta}_{\phantom{\mu\nu;\beta};\beta}                     
 -R^{\mu\beta;\nu}_{\phantom{\mu\beta;\nu};\beta}                        
-R^{\nu \beta;\mu}_{\phantom{\nu \beta;\mu};\beta}                          
 - 2R^{\mu\beta}R^{\nu}_{\phantom{\nu}\beta}                                    
+\frac{1}{2}g^{\mu\nu}R_{\alpha\beta}R^{\alpha\beta}.
\label{E47}
\end{equation}                                 

As a classical theory conformal gravity is of interest  since, as can be seen from the structure of (\ref{E46}) and (\ref{E47}),  the Ricci-flat Schwarschild metric is an exact vacuum solution to it. In principle then, one can thus bypass the Einstein-Hilbert action altogether and still recover the standard solar system phenomenology. Indeed, as had been noted in \cite{R4}, for a gravity theory to be viable, it only needs to recover the solutions to Einstein gravity in the kinematic region in which they have tested, and one does not at all need to recover the Einstein equations themselves. Alternate theories whose solutions reduce to the Schwarzschild solution on solar system distance scales can thus satisfy the three classic tests of Einstein gravity even if they never reduce to the Einstein equations themselves. Moreover, while the vanishing of $W_{\mu\nu}$ is satisfied by $R_{\mu\nu}=0$, since $W_{\mu\nu}$ is a derivative function of $R_{\mu\nu}$, the vacuum vanishing of $W_{\mu\nu}$ can be achieved without $R_{\mu\nu}$ needing to vanish. Conformal gravity thus has other, non-Schwarzschild, solutions as well, with the general solution found by Mannheim and Kazanas \cite{R11} replacing the metric coefficients of the Schwarzschild solution by metric coefficients of the form $-g_{00}=1/g_{rr}=1-2\beta/r+\gamma r-kr^2$ where $\beta$ and $\gamma$ are (source-dependent) constants. In this solution the $\gamma r$ and $-kr^2$ terms only lead to departures from Schwarzschild at large distances but not at small ones, with detailed analysis \cite{R4,R11a} showing that the conformal theory is able to provide for an accounting of the observed systematics of galactic rotation curves without the need for any of the dark matter that is required in the standard theory. 

Now while the interest of the present paper is in the quantum aspects of conformal gravity, for a quantum theory to be viable it would still have to be able to recover standard solar system physics. It is because the Einstein theory does contain the standard solar system phenomenology, that it is thought that it provides the correct, and even the only possible,  starting point for constructing a quantum gravity theory. However, as the above discussion shows, Einstein is only sufficient to give Schwarschild but not necessary. It is thus legitimate to base a quantum gravity theory on some non-Einstein starting point, as long as the alternate theory also possesses the Schwarzschild phenomenology. One should thus retain the kinematic aspects of general relativity (metric nature of gravity, general covariance, equivalence principle), but consider the possibility that dynamically one might need to change the equations of motion that the metric is to obey, with conformal gravity suggesting that the change is to be in replacing second-order derivative gravitational equations by fourth-order ones.

Thus motivated, we now proceed to quantize the conformal theory, and for our purposes here it will suffice to linearize the theory around flat spacetime to second order in the metric fluctuations. Given the experience learned above in constructing a quantum theory in 2D, our first concern in quantizing conformal gravity is the ordering issue, and especially so since at the classical level the Lanczos Lagrangian $L_{\rm L}$ given in (\ref{E43}) is a total derivative. However, it turns out that, to second quantum order at least, even after allowing for ordering, no commutator terms are encountered, with $L_{\rm L}$ remaining a total derivative. Specifically, unlike the Einstein-Hilbert action, the Weyl action is second order in the Riemann curvature rather than first. Consequently, when linearized around flat spacetime, one only needs to expand the Riemann and Ricci tensor terms in $I_{\rm W}$ to first order. However, in the Einstein-Hilbert action case, the problematic terms arose from the Christoffel symbol product term $g_{\eta\sigma}(\Gamma^{\eta}_{\nu\lambda}\Gamma^{\sigma}_{\mu\kappa}-\Gamma^{\eta}_{\kappa\lambda}\Gamma^{\sigma}_{\mu\nu})$ in the Riemann tensor as given in (\ref{E1}). Since this product term is second order in the fluctuation around flat spacetime, it does not appear in the Weyl action to the order of interest, with everything in the action being obtainable from the $(1/2)\left(\partial_{\kappa}\partial_{\mu} g_{\lambda\nu}  -\partial_{\kappa}\partial_{\lambda} g_{\mu\nu}  -\partial_{\nu}\partial_{\mu}g_{\lambda\kappa} +\partial_{\nu}\partial_{\lambda}g_{\mu\kappa} \right)$ term in the Riemann tensor alone, Consequently, to the quantum-mechanical order of interest to us here, we can use (\ref{E45}) as is.

Unlike the 2D Einstein case, this time the equation of motion $W^{\mu\nu}(1)=0$ will have both trivial and non-trivial solutions, and with ordering not being an issue, the linearized quantum-mechanical $W^{\mu\nu}_{\rm M}(2)$ will be covariantly conserved in the non-trivial ones, just as it would be classically. In this respect, the key role of the coupling to a quantum source is to force gravity to actually have to choose the non-trivial solution to the equation $W^{\mu\nu}_{\rm M}(1)=0$. It is thus the non-vanishing of the quantum $T^{\mu\nu}_{\rm M}$ that forces $W^{\mu\nu}(2)$, and thus $h_{\mu\nu}$, to be non-zero, with the covariant conservation of $T^{\mu\nu}_{\rm M}$ forcing the covariant conservation of $W^{\mu\nu}(2)$ and thus the non-trivial vanishing of  $W^{\mu\nu}(1)$, rather than the other way round \cite{R12}. 

On now explicitly linearizing  $W^{\mu\nu}$ of (\ref{E45}), to first order in $h_{\mu\nu}$ we obtain
\begin{equation}
W^{\mu\nu}(1)=\frac{1}{2}\Pi^{\mu\rho}
\Pi ^{\nu\sigma}K_{\rho \sigma}- \frac{1}{6}\Pi^{\mu \nu} \Pi ^{\rho
\sigma}K_{\rho\sigma},
\label{E48}
\end{equation}
where
\begin{equation}
K^{\mu
\nu}=h^{\mu \nu}-
\frac{1}{4}\eta^{\mu \nu} h^{\alpha}_{\phantom{\alpha}\alpha},\qquad
\Pi^{\mu \nu}
=\eta^{\mu \nu} \partial^{\alpha}\partial_{\alpha}-
\partial^{\mu}\partial^{\nu}.
\label{E49}
\end{equation}
Since choosing the non-trivial solution to $W^{\mu\nu}(1)=0$ is going to be forced upon us by the second order $-4\alpha_g W^{\mu\nu}+T^{\mu\nu}_{\rm M}=0$, it is instructive to  analyze the solutions to $W^{\mu\nu}(1)=0$ right away as it will enable us to identify a very convenient gauge in which to calculate $W^{\mu\nu}(2)$. In (\ref{E48}) we note that only the traceless combination $K^{\mu\nu}=h^{\mu \nu}-(1/4) \eta^{\mu \nu} h^{\alpha}_{\phantom{\alpha}\alpha}$ appears, with only nine components of $h_{\mu\nu}$ being of relevance in the conformal case. Since there are four gauge transformations of the form $h_{\mu\nu}\rightarrow h_{\mu\nu}+\partial_{\mu}\epsilon_{\nu}+\partial_{\nu}\epsilon_{\mu}$ at our disposal, we can reduce $K_{\mu\nu}$ to five independent components. The most convenient gauge choice is make $K_{\mu\nu}$ be transverse and obey $\partial_{\mu}K^{\mu\nu}=0$. We can thus take the general $K_{\mu\nu}$ to be transverse traceless \cite{R13}. 

The utility of having a transverse-traceless $K_{\mu\nu}$  is that with it (\ref{E48}) simplifies to 
\begin{equation}
W^{\mu\nu}(1)=\frac{1}{2}(\partial_{\alpha}\partial^{\alpha})^2 K^{\mu\nu},
\label{E50}
\end{equation}
with there now being no mixing of components of $K_{\mu\nu}$. By the same token, in this same gauge the second order contribution to $I_{\rm W}$ reduces to 
\begin{equation}
I_{\rm W}=-\frac{\alpha_g}{2}\int d^4x \partial_{\alpha}\partial^{\alpha} K_{\mu\nu}\partial_{\beta}\partial^{\beta} K^{\mu\nu},
\label{E51}
\end{equation}
with $I_{\rm W}$ equally possessing no mixing of components of $K_{\mu\nu}$ through second order. In consequence of this, $W^{\mu\nu}(2)$ will also only depend on $K_{\mu\nu}$ and not on the trace $h$ of the fluctuation, and it too will greatly simplify. Similarly, when $W^{\mu\nu}(1)$ is zero, in a  linearization around flat spacetime the condition  $[\nabla_{\mu}W^{\mu\nu}](2)=0$ simplifies to $\partial_{\mu}W^{\mu\nu}(2)=0$.

Since (\ref{E50}) involves the fourth-order wave operator, the wave equation $W^{\mu\nu}(1)=0$ will have twice as many independent solutions as the usual second-order wave equation. One set of solutions will be those that do satisfy the second-order wave equation (since they then automatically satisfy the fourth-order wave equation as well), with the other set being modes that have the property that after being acted on by the second-order wave operator, one obtains functions that then satisfy the usual second-order wave equation. In terms of all of these basis modes the most general solution to $W^{\mu\nu}(1)=0$ can be written as 
\begin{equation}
K_{\mu\nu}=A_{\mu\nu}e^{ik\cdot x}+B_{\mu\nu}(n\cdot x)e^{ik\cdot x}
\label{E52}
\end{equation}
where $k^{\mu}$ is a lightlike 4-vector $k^{\mu}=(\omega_k,\bar{k})$ with $\omega_k=|\bar{k}|$, $n^{\mu}=(1,0,0,0)$ is a unit timelike vector that obeys $n^{\mu}n_{\mu}=-1$, $n^{\mu}x_{\mu}=-t$, and $A_{\mu\nu}$ and $B_{\mu\nu}$ are traceless polarization tensors. (We shall comment on the origin of the timelike 4-vector $n^{\mu}$ below in Sec. \ref{S8}.)

The requirement that the $A_{\mu\nu}e^{ik\cdot x}$ modes be transverse obliges $A_{\mu\nu}$ to obey $k_{\mu}A^{\mu\nu}=0$, while the requirement that the $B_{\mu\nu}e^{ik\cdot x}$ modes be transverse obliges $B_{\mu\nu}$ to obey $k_{\mu}B^{\mu\nu}=0$, $n_{\mu}B^{\mu\nu}=0$. The conditions on $B_{\mu\nu}$ reduce it to two independent components. For instance, for the typical lightlike vector $k^{\mu}=(\omega_k,0,0,k)$, the only non-zero components of $B_{\mu\nu}$ are $B_{11}=-B_{22}$ and $B_{12}$. The condition imposed on $A_{\mu\nu}$ reduces it to five independent components, but we can reduce $A_{\mu\nu}$ further to only two components via residual gauge transformations that preserve the wave equation and leave the basis modes transverse and traceless. For the general $K_{\mu\nu}$, a general gauge transformation on $h_{\mu\nu}$ effects $K_{\mu\nu}\rightarrow K_{\mu\nu}+\partial_{\mu}\epsilon_{\nu}+\partial_{\nu}\epsilon_{\mu}-(1/2)\eta_{\mu\nu}\partial_{\alpha}\epsilon^{\alpha}$. It automatically leaves $K_{\mu\nu}$ traceless, and will leave it transverse if $\epsilon_{\mu}$ obeys $\partial_{\alpha}\partial^{\alpha}\epsilon_{\nu}+(1/2)\partial_{\nu}\partial^{\alpha}\epsilon_{\alpha}=0$. Taking $\epsilon_{\mu}$ to be of the form $f_{\mu}e^{ik\cdot x}$ with lightlike  $k_{\mu}$, the residual gauge transformations are thus required to obey $k_{\mu}f^{\mu}=0$, to thereby give three independent $f_{\mu}$ and enable us to reduce $A_{\mu\nu}$ to two independent components. For the lightlike vector $k^{\mu}=(\omega_k,0,0,k)$ for instance, the choice $f^{0}=-A^{00}/2i\omega_k$,  $f^{1}=-A^{01}/i\omega_k$, $f^{2}=-A^{02}/i\omega_k$ brings $A_{\mu\nu}$ to a form in which its only non-zero components are $A_{11}=-A_{22}$ and $A_{12}$, to thus bring $A_{\mu\nu}$ and $B_{\mu\nu}$ to the same generic form. To separate out the two polarization states, for 4-vectors such as the vector $k^{\mu}=(\omega_k,0,0,k)$ we shall let $\epsilon_{\mu\nu}^{(1)}(\bar{k})$ denote the polarization tensor with $\epsilon_{11}=-\epsilon_{22}=1/\surd{2}$,  and shall let $\epsilon_{\mu\nu}^{(2)}(\bar{k})$ denote the polarization tensor with $\epsilon_{12}=\epsilon_{21}=1/\surd{2}$ (both as normalized to $\epsilon_{\alpha\beta}\epsilon^{\alpha\beta}=1$).

To allow for the possibility that the negative frequency operators may not be the Hermitian conjugates of the positive frequency ones (as will in fact turn out to be the case), guided by the coupling-constant structure given in (\ref{E15}), and using a hat notation to denote the creation operators, we take the general $K_{\mu\nu}$ operator to be of the form 
\begin{eqnarray}
K_{\mu\nu}(x)&=&\frac{\hbar^{1/2}}{2(-\alpha_g)^{1/2}}\sum_i\int \frac{d^3k}{(2\pi)^{3/2}(\omega_k)^{3/2}}\bigg{[}
A^{(i)}(\bar{k})\epsilon^{(i)}_{\mu\nu}(\bar{k})e^{ik\cdot x}
+i\omega_kB^{(i)}(\bar{k})\epsilon^{(i)}_{\mu\nu}(\bar{k})(n\cdot x)e^{ik\cdot x}\
\nonumber\\
&+&\hat{A}^{(i)}(\bar{k})\epsilon^{(i)}_{\mu\nu}(\bar{k})e^{-ik\cdot x}-
i\omega_k\hat{B}^{(i)}(\bar{k})\epsilon^{(i)}_{\mu\nu}(\bar{k})(n\cdot x)e^{-ik\cdot x}\bigg{]},
\label{E53}
\end{eqnarray}
where $\hat{A}^{(i)}(\bar{k})\epsilon^{(i)}_{\mu\nu}(\bar{k})$ and $\hat{B}^{(i)}(\bar{k})\epsilon^{(i)}_{\mu\nu}(\bar{k})$ have the same generic polarization tensor structure as $A^{(i)}(\bar{k})\epsilon^{(i)}_{\mu\nu}(\bar{k})$ and $B^{(i)}(\bar{k})\epsilon^{(i)}_{\mu\nu}(\bar{k})$.

For zero-point energy density purposes, we only need to determine the vacuum expectation value of $W_{\mu\nu}(2)$. On taking $A^{(i)}(\bar{k})$ and $B^{(i)}(\bar{k})$ to annihilate the right vacuum and $\hat{A}^{(i)}(\bar{k})$ and $\hat{B}^{(i)}(\bar{k})$ to annihilate the left vacuum as usual, and on taking the commutators of the positive frequency $A^{(i)}(\bar{k})$ and $B^{(i)}(\bar{k})$ operators with the negative frequency $\hat{A}^{(j)}(\bar{k}^{\prime})$ and $\hat{B}^{(j)}(\bar{k}^{\prime})$ operators to behave as $\delta^3(\bar{k}-\bar{k}^{\prime})\delta_{i,j}$, then following a lengthy but straightforward calculation,  we find that $-4\alpha_g\langle \Omega|W_{\mu\nu}(2)|\Omega \rangle$ evaluates to
\begin{equation}
-4\alpha_g\langle \Omega|W_{\mu\nu}(2)|\Omega \rangle=\hbar\int \frac{d^3k}{(2\pi)^3\omega_k}\langle \Omega|X_{\mu\nu}(k)|\Omega \rangle,
\label{E54}
\end{equation}
where the operator $X_{\mu\nu}(k)$ is given by
\begin{eqnarray}
X_{\mu\nu}(k)&=&\sum_i\bigg{[}\left[\frac{1}{2}k_{\mu}k_{\nu}(n \cdot n)-(k_{\mu}n_{\nu}+k_{\nu}n_{\mu})(k\cdot n)
+\frac{1}{2}\eta_{\mu\nu}(k\cdot n)^2\right](\hat{B}^{(i)}B^{(i)}+B^{(i)}\hat{B}^{(i)})
\nonumber\\
&-&\frac{k_{\mu}k_{\nu}}{\omega_k}(k\cdot n)\left[\hat{A}^{(i)}B^{(i)}+A^{(i)}\hat{B}^{(i)}]
+ik_{\mu}k_{\nu}(k\cdot n)(\hat{B}^{(i)}B^{(i)} -B^{(i)}\hat{B}^{(i)})(n \cdot x)\right]
\bigg{]}.
\nonumber\\
\label{E55}
\end{eqnarray}
As constructed the trace $\eta^{\mu\nu}X_{\mu\nu}(k)$ is manifestly zero, just as it should be. While we find the presence of a time-dependent $(n \cdot x)$ term in (\ref{E55}), we can cancel it by setting
\begin{equation}
\hat{B}^{(1)}(\bar{k})B^{(1)}(\bar{k})-B^{(1)}(\bar{k})\hat{B}^{(1)}(\bar{k})=0,\qquad \hat{B}^{(2)}(\bar{k})B^{(2)}(\bar{k})-B^{(2)}(\bar{k})\hat{B}^{(2)}(\bar{k})=0.
\label{E56}
\end{equation}
While this is not a conventional commutation requirement on creation and annihilation operators, an analog of it was already encountered in \cite{R1,R2}, and we defer discussion of this point to Sec. \ref{S8}, as it is characteristic of a theory whose Hamiltonian is of a non-diagonalizable, and thus non-Hermitian, Jordan-block form. With this requirement  $X_{\mu\nu}(k)$ reduces to
\begin{eqnarray}
X_{\mu\nu}(k)&=&\sum_i\left[k_{\mu}k_{\nu}(n \cdot n)-2(k_{\mu}n_{\nu}+k_{\nu}n_{\mu})(k\cdot n)
+\eta_{\mu\nu}(k\cdot n)^2\right]\hat{B}^{(i)}(\bar{k})B^{(i)}(\bar{k})
\nonumber\\
&-&\frac{k_{\mu}k_{\nu}}{\omega_k}(k\cdot n)\left[\hat{A}^{(i)}(\bar{k})B^{(i)}(\bar{k})+A^{(i)}(\bar{k})\hat{B}^{(i)}(\bar{k})
\right].
\label{E57}
\end{eqnarray}

Having obtained a closed form expression for $-4\alpha_g\langle \Omega|W_{\mu\nu}(2)|\Omega \rangle$, we can proceed to see how it can cancel against the vacuum expectation value of the matter field energy-momentum tensor. However, in order to do so, we first need to calculate the spatial integral of the operator $-4\alpha_gW_{00}(2)$ as it serves as the Hamiltonian operator of the gravitational field, and it is found to take the form
\begin{eqnarray}
&&-4\alpha_g\int d^3 x W_{00}(2)
\nonumber\\
&&\qquad =\sum_i\int d^3k\hbar\omega_k\left[\hat{A}^{(i)}(\bar{k})B^{(i)}(\bar{k})+A^{(i)}(\bar{k})\hat{B}^{(i)}(\bar{k}))+2\hat{B}^{(i)}(\bar{k})B^{(i)}(\bar{k})\right],
\label{E58}
\end{eqnarray}
where $\omega_k=|k|$. It is from (\ref{E58}) that we shall identify the Hilbert space and commutator algebra that is to be associated with the gravitational field, and to actually make the identification we turn to an analysis of the ghost problem in fourth-order derivative theories and its resolution.

\subsection{Ghost Problem of Second- Plus Fourth-Order Theories}
\label{S7}

In order to analyze the quantum structure of pure fourth-order derivative theories, it is very helpful to first add on to the theory a second-order derivative term, and then track the limit in which the second-order term is subsequently switched off. Since the transverse gauge, linearized conformal gravity action given in (\ref{E51}) is diagonal in its $K_{\mu\nu}$ components, we can ignore the $(\mu\nu)$ indices and consider a scalar field theory. As well as the simplification that this provides, by dropping the spacetime indices, we are able to differentiate between ghost problems associated with the Hilbert space metric and ghost problems that are associated with the negative signature of the spacetime metric. Hilbert space metric issues are met in higher-derivative field theories even when no gauge fields are involved at all as they are already present in higher-derivative scalar field theories. Spacetime metric signature issues can be met when continuous symmetries are given a local extension. In this paper we deal only with the conformal gravity Hilbert space metric issues, and bypass spacetime metric issues by working throughout in a single gauge, one in which the metric fluctuations are conveniently taken to be transverse.

For a higher-derivative flat spacetime  scalar field theory, we take as action the action $I_{\rm S}$ given in \cite{R2}, viz.
\begin{equation}
I_{\rm S}=-\frac{1}{2}\int d^4x\left[\partial_{\mu}\partial_{\nu}\phi\partial^{\mu}
\partial^{\nu}\phi+(M_1^2+M_2^2)\partial_{\mu}\phi\partial^{\mu}\phi+M_1^2M^2_2\phi^2\right],
\label{E59}
\end{equation}
where $M_1$ and $M_2$ are in wave number units.
The variation of this action gives an equation of motion of the form
\begin{equation}
(-\partial_t^2+\bar{\nabla}^2-M_1^2)(-\partial_t^2+\bar{\nabla}^2-M_2^2)\phi(x)=0.
\label{E60}
\end{equation}
With $k^2=-k_0^2+\bar{k}^2$, the propagator of the theory is given by
\begin{eqnarray}
D^{(4)}(x,x^{\prime},M_1,M_2)&=&\int \frac{d^4k}{(2\pi)^4}\frac{e^{ik\cdot (x-x^{\prime})}}{(k^2+M_1^2)(k^2+M_2^2)}
\nonumber \\
&=&\int \frac{d^4k}{(2\pi)^4}\frac{e^{ik\cdot (x-x^{\prime})}}{(M_2^2-M_1^2)}\left(\frac{1}{k^2+M_1^2}-\frac{1}{k^2+M_2^2}\right).
\label{E61}
\end{eqnarray}

As constructed, the immediate appeal of the propagator of (\ref{E61}) is that it has very good $1/k^4$ convergence in the ultraviolet rather than the $1/k^2$ behavior found in second-order theories, and it is for this reason that fourth-order derivative theories of gravity can be renormalizable, even while  second-order derivative theories of gravity are not. However, this good ultraviolet convergence is thought to come at a price, namely the possible presence of negative norm ghost states and a concomitant violation of unitarity. Specifically, with a Feynman frequency contour integration giving  a propagator of the form
\begin{eqnarray}
D^{(4)}(\bar{x},\bar{x}^{\prime},t,M_1,M_2)&=&\frac{1}{(M_1^2-M_2^2)}\int \frac{d^3k}{(2\pi)^3}\frac{e^{i\bar{k}\cdot (\bar{x}-\bar{x}^{\prime})}}{2\omega_k^1}\left[\theta(t)e^{-i\omega_k^1t}+\theta(-t)e^{i\omega_k^1t} \right]
\nonumber\\
&-&\frac{1}{(M_1^2-M_2^2)}\int \frac{d^3k}{(2\pi)^3}\frac{e^{i\bar{k}\cdot (\bar{x}-\bar{x}^{\prime})}}{2\omega_k^2}\left[\theta(t)e^{-i\omega_k^2t}+\theta(-t)e^{i\omega_k^2t} \right]
\label{E62}
\end{eqnarray}
($\omega_k^1=(\bar{k}^2+M_1^2)^{1/2}$, $\omega_k^2=(\bar{k}^2+M_2^2)^{1/2}$), one anticipates that the completeness relation for the states will be given by the indefinite metric relation
\begin{equation}
\sum_n |n\rangle\langle n|- \sum_m |m\rangle\langle m|=1,
\label{E63}
\end{equation}
to thus involve states of negative Dirac norm.

However, it turns out that on actually quantizing the theory and constructing the appropriate Hilbert space \cite{R1,R2}, this anticipation turns out to be incorrect. In hindsight, that it must be incorrect can actually be inferred from the structure of (\ref{E61}). Specifically, since the propagator in (\ref{E61}) is in the form of a difference of two perfectly normal second-order propagators, viz.
\begin{equation}
D^{(4)}(\bar{x},\bar{x}^{\prime},t,M_1,M_2)=\frac{1}{(M_2^2-M_1^2)}\left(D^{(2)}(\bar{x},\bar{x}^{\prime},t,M_1)-D^{(2)}(\bar{x},\bar{x}^{\prime},t,M_2)\right),
\label{E64}
\end{equation}
we see that it consists of two components each one of which has a set of basis modes that is complete and has energies that are all real (all of the poles in (\ref{E61}) lie on the real frequency axis). Consequently, the modes of the fourth-order propagator are also complete (i.e. double the basis of a single second-order propagator) and have real energies. Now any Hamiltonian whose eigenspectrum is real and complete is either Hermitian already or can be brought to a Hermitian form by a similarity transform. Since the completeness relation given in (\ref{E63}) is to hold in the energy eigenmode basis, and since one can always take the Hilbert space of a Hermitian matrix to have a positive definite metric (i.e. time evolution with a Hermitian Hamiltonian is always unitary), it must be the case that (\ref{E63}) cannot hold and that the Hamiltonian of the fourth-order theory must instead not be Hermitian. 

In the actual study of \cite{R1,R2} this lack of Hermiticity was traced to the fact that on expressly constructing the quantum Hamiltonian as a differential operator in phase space, the wave functions of the eigenvectors were found to not be normalizable on the real axis. Consequently, on the real axis one could not represent all of the momentum operators as $-i\hbar \partial_x$ type derivatives, with some of the momentum operators not being Hermitian, and the Hamiltonian built out of them then not being Hermitian either \cite{R14}. Moreover, in \cite{R1,R2} the similarity transformation that then brings the Hamiltonian to a Hermitian form was expressly constructed, to thus show that an associated Hermitian form does indeed exist.

Since the Hamiltonian can be related to a Hermitian Hamiltonian by a similarity transformation, it must equally lead to unitary time evolution, and the theory must be unitary after all. Now we are accustomed in quantum mechanics to using the standard Dirac norm in which the energy eigenbras are the conjugates of the energy eigenkets. However, such an association only holds for Hermitian Hamiltonians, since for real  energy eigenvalues and Hermitian Hamiltonians, from $H|\psi\rangle =E|\psi \rangle$ it follows that $\langle \psi|H=\langle \psi|E$. However, if the Hamiltonian is not Hermitian but still has real eigenvalues (Hermiticity is only sufficient to give real eigenvalues but not necessary) this time from $H|\psi\rangle =E|\psi \rangle$ it follows that $\langle \psi|H^{\dagger}=\langle \psi|E$, with the conjugate of the eigenket no longer being an energy eigenstate of $H$. Thus for non-Hermitian Hamiltonians with real eigenvalues one must distinguish between left- and right-eigenvectors according to 
\begin{equation}
H|R\rangle =E|R \rangle,\qquad \langle L|H=\langle L|E,
\label{E65}
\end{equation}
with $\langle L|$ not being the same as $\langle R|$. In the non-Hermitian case it is $\langle L|R \rangle$ rather than $\langle R|R \rangle$ that is the appropriate inner product for the Hilbert space, with the $\langle L(t)|R(t) \rangle=\langle L(t=0)|e^{iHt/\hbar}e^{-iHt/\hbar}|R(t=0) \rangle=\langle L(t=0)|R(t=0) \rangle$ norm being preserved in time because of (\ref{E65}), even as the $\langle R(t)|R(t) \rangle=\langle R(t=0)|e^{iH^{\dagger}t/\hbar}e^{-iHt/\hbar}|R(t=0) \rangle$ norm is not. 

As we see, for non-Hermitian Hamiltonians with real eigenvalues, it is the $\langle L|R \rangle$ norm that one must use, and if in a theory one finds the standard $\langle R|R \rangle$ norm to be negative, this does not necessarily mean that the theory is sick. It could simply mean that the Hamiltonian is not Hermitian and that one is not in the correct Hilbert space for the theory, with the presence of Dirac norm ghost states being a diagnostic rather than a disease \cite{R15}. While the left-eigenvectors of a non-Hermitian Hamiltonian with real eigenvalues are not the conjugates of the right-eigenvectors, they can be related to them by a similarity transform $\langle L|=\langle R|S$, with the appropriate norm for the Hilbert space thus being given as $\langle R|S|R \rangle$ rather than as $\langle R|R \rangle$ itself. Moreover, this same operator $S$ is related to the similarity transformation that brings a non-Hermitian matrix  with a real and complete eigenspectrum to Hermitian form. (If we set $S=e^{-\cal{Q}}$, then $\tilde{H}=e^{-{\cal{Q}}/2}He^{{\cal{Q}}/2}$ is Hermitian \cite{R9}.) Thus, rather than represent the propagator of (\ref{E61}) as the Green's function $\langle \Omega_R|T(\phi(x)\phi(x^{\prime}))|\Omega_R\rangle$, we must instead represent it as $\langle \Omega_L|T(\phi(x)\phi(x^{\prime}))|\Omega_R\rangle=\langle \Omega_R|ST(\phi(x)\phi(x^{\prime}))|\Omega_R\rangle$, with the bra $\langle \Omega_L|$ that the negative-frequency component of $\phi(x)$ left-annihilates not being the conjugate of the ket $|\Omega_R\rangle$ that the positive-frequency component of $\phi(x)$ right-annihilates. It is the presence of the $S$ operator in the Green's function, and not the existence of negative norm states, that is the origin of the relative minus sign that appears in (\ref{E61}). Specifically, as noted in \cite{R2}, the relative minus sign explicitly originates as the $-1$ eigenvalue of an operator of the theory, the so-called $\cal{C}$ operator of $\cal{P}\cal{T}$ theories \cite{R9} that obeys ${\cal{C}}^2=1$ and $[{\cal{C}},H]=0$, an operator that is related to $S$ according to $S=\cal{P}\cal{C}$. (Here $\cal{P}$ and $\cal{T}$ are the parity and time reversal operators, or equivalently, some appropriate generalization of them in which one operator is discrete and linear and the other is discrete and anti-linear.) 

One of the most interesting aspects of the  $\langle L|R \rangle=\langle R|S|R \rangle$ norm is that it depends on $S$. However $S$ itself is not universal. Rather, it depends on the Hamiltonian of interest as it is constructed from an operator $\cal{C}$ that is required to commute with the Hamiltonian. (For theories in which the Hamiltonian commutes with the $\cal{P}\cal{T}$ product, it can be shown that a non-trivial $\cal{C}$ operator will always exist \cite{R16}.) The $\langle R|S|R \rangle$ norm is thus dynamically determined, with each non-Hermitian Hamiltonian with a real eigenspectrum having its own appropriate inner product. This situation is to be contrasted with the standard Dirac norm, as it is assigned independent of dynamics and can be used for any Hamiltonian that is Hermitian. As such, this situation is analogous to the difference between special and general relativity, since special relativity only requires a pre-assigned, dynamics-independent spacetime Minkowski metric, while in general relativity the metric is determined by the dynamics itself. Thus the error made in anticipating that the fourth-order propagator of (\ref{E61}) involves ghost states is in thinking that in quantum mechanics the Hilbert space inner product can always be pre-assigned independent of the choice of Hamiltonian. While this is correct for Hermitian Hamiltonians, it is not correct for non-Hermitian ones, with the Hilbert space inner product then having to be constructed on a case by case basis, with such theories then being able to be unitary with respect to the appropriate norm.

Now we had noted above that despite the fact that the Hamiltonian associated with the fourth-order scalar field theory is not Hermitian, the energy eigenvalues are nonetheless real. Thus we have to ask why this is the case. As noted in \cite{R1,R2} the reason for this is that while it is not Hermitian, the Hamiltonian falls into the class of $\cal{P}\cal{T}$-invariant Hamiltonians that have real eigenvalues (see \cite{R9} for a general review of such $\cal{P}\cal{T}$ theories). Specifically, it had been noted in \cite{R17} that for Hamiltonians that obey $[H,{\cal{P}\cal{T}}]=0$, while the energy eigenvalues would not necessarily be real ($\cal{T}$ is anti-linear rather than linear), the secular equation that determines the eigenvalues would be, with the eigenvalues then either being real or appearing in complex conjugate pairs. More recently, the converse was shown \cite{R16}, namely, that if the secular equation is real, the Hamiltonian must commute with $\cal{P}\cal{T}$. In consequence, if a Hamiltonian is not $\cal{P}\cal{T}$ invariant, not all of its eigenvalues can be real. Since the eigenvalues of the fourth-order scalar field theory with action $I_{\rm S}$ are all real, its Hamiltonian must be $\cal{P}\cal{T}$ invariant. Moreover, it was also shown in \cite{R16}, that the eigenvalues of a $\cal{P}\cal{T}$-invariant Hamiltonian will all be real unless there exists a $\cal{C}$ operator that does not commute with $\cal{P}\cal{T}$. For the fourth-order scalar field theory there is no such $\cal{C}$ operator, and all of its energy eigenvalues thus have to be real, just as required. In this case the $\langle L|R \rangle$ norm coincides with the $\cal{C}\cal{P}\cal{T}$ norm \cite{R9} that had been developed for $\cal{P}\cal{T}$ theories \cite{R18}.

To explicitly construct the Hamiltonian associated with the $I_{\rm S}$ action, one can proceed canonically \cite{R19} and construct the energy-momentum tensor and the canonical momenta using a procedure that was developed by Ostrogradski in the nineteenth century to handle Lagrangians with higher derivatives \cite{R20}, or equivalently one can use the method of Dirac constraints as applied to fourth-order theories in \cite{R21}. As discussed in \cite{R2}, for the action $I_{\rm S}$ with Lagrange density $\cal{L}$, the Ostrogradski  procedure leads to canonical conjugates and an energy-momentum tensor of the form ($\phi_{,\mu}=\partial_{\mu}\phi=\partial \phi/\partial x^{\mu}$)
\begin{equation}
\pi^{\mu}=\frac{\partial{\cal L}}{\partial \phi_{,\mu}}-\partial_{\lambda
}\left(\frac{\partial {\cal L}}{\partial\phi_{,\mu,\lambda}}\right)=-(M_1^2+M_2^2)\partial^{\mu}\phi+\partial_{\lambda}\partial^{\mu}\partial^{\lambda}\phi,
\qquad \pi^{\mu\lambda}=\frac{\partial {\cal L}}{\partial \phi_{,\mu,\lambda}}=-\partial^{\mu}\partial^{\lambda}\phi,
\label{E66}
\end{equation}
\begin{eqnarray}
T^{\rm OST}_{\mu\nu}(M_1,M_2)&=&\pi_{\mu}\phi_{,\nu}+\pi_{\mu\lambda}\phi^{,\lambda}_{\phantom{\lambda},\nu}-\eta_{\mu\nu}{\cal L}
\nonumber\\
&=&\pi_{\mu}\phi_{,\nu}-\pi_{\mu\lambda}\pi^{\lambda}_{\phantom{\lambda}
\nu}+\frac{1}{2}\eta_{\mu\nu}\left[\pi_{\lambda\kappa}\pi^{\lambda\kappa}+(M_1^2
+M_2^2)\partial_{\lambda}\phi\partial^{\lambda}\phi+M_1^2M_2^2\phi^2\right],
\nonumber\\
T^{\rm OST}_{00}(M_1,M_2)&=&
\pi_{0}\dot{\phi}+\frac{1}{2}\left[\pi_{00}^2+(M_1^2
+M_2^2)(\dot{\phi}^2-\partial_{i}\phi\partial^{i}\phi)-M_1^2M_2^2\phi^2
-\pi_{ij}\pi^{ij}\right].
\label{E67}
\end{eqnarray}
With $T^{\rm OST}_{\mu\nu}(M_1,M_2)$ being covariantly conserved, in solutions to the equation of motion (\ref{E60}) the Hamiltonian $H^{\rm OST}(M_1,M_2)=\int d^3x T^{\rm OST}_{00}(M_1,M_2)$ is time independent.

In (\ref{E66}) we recognize two canonical momenta $\pi_{0}$ and $\pi^{0}_{\phantom{0}0}$, which respectively serve as canonical conjugates to $\phi$ and $\partial_{0}\phi$ according to $[\phi(\bar{x},t),\pi_0(\bar{x}^{\prime},t)]=i\hbar\delta^3(\bar{x}-\bar{x}^{\prime})$, $[\partial_0\phi(\bar{x},t),\pi^{0}_{\phantom{0}0}(\bar{x}^{\prime},t)]=i\hbar\delta^3(\bar{x}-\bar{x}^{\prime})$. (In the fourth-order case $\phi(\bar{x},t)$ and $\partial_0\phi(\bar{x}^{\prime},t)$ are not conjugates and the $[\phi(\bar{x},t),\partial_0\phi(\bar{x}^{\prime},t)]$ equal-time commutator is zero.) Because the Hamiltonian and the two canonical momenta are not Hermitian, one has to make an expansion of the field operators in terms of two sets of creation and annihilation operators that are not in the form of Hermitian conjugate pairs. And since these operators are not Hermitian conjugate pairs, we use a hat notation to denote the creation operators.  Following \cite{R2}, we expand the scalar field as
 \begin{eqnarray}
\phi(x)&=&\hbar^{1/2}\int \frac{d^3k}{(2\pi)^{3/2}(2\omega^1_k)^{1/2}}[a_{1,\bar{k}}e^{i(\bar{k}\cdot \bar{x} -\omega^1_kt)} 
+\hat{a}_{1,\bar{k}}e^{-i(\bar{k}\cdot \bar{x} -\omega^1_kt)}]
\nonumber \\
&+&\hbar^{1/2}\int \frac{d^3k}{(2\pi)^{3/2}(2\omega^2_k)^{1/2}}[ia_{2,\bar{k}}e^{i(\bar{k}\cdot \bar{x} -\omega^2_kt)} 
+i\hat{a}_{2,\bar{k}}e^{-i(\bar{k}\cdot \bar{x} -\omega^2_kt)}]
\label{E68}
\end{eqnarray}
where $\omega^1_k=(\bar{k}^2+M_1^2)^{1/2}$,  $\omega^2_k=(\bar{k}^2+M_2^2)^{1/2}$. On constructing the canonical commutation relations, one obtains the algebra 
\begin{eqnarray}
&&[a_{1,\bar{k}},\hat{a}_{1,\bar{k}^{\prime}}]=\frac{1}{(M_1^2-M_2^2)}\delta^3(\bar{k}-\bar{k}^{\prime}),
\qquad [a_{2,\bar{k}},\hat{a}_{2,\bar{k}^{\prime}}]=\frac{1}{(M_1^2-M_2^2)}\delta^3(\bar{k}-\bar{k}^{\prime}),
\nonumber\\
&&[a_{1,\bar{k}},a_{2,\bar{k}^{\prime}}]=0,\quad[a_{1,\bar{k}},\hat{a}_{2,
\bar{k}^{\prime}}]=0,\quad[\hat{a}_{1,\bar{k}},a_{2,\bar{k}^{\prime}}]=0,\quad
[\hat{a}_{1,\bar{k}},\hat{a}_{2,\bar{k}^{\prime}}]=0,
\label{E69}
\end{eqnarray}
with the Hamiltonian evaluating to the diagonal form
 \begin{equation}
H^{\rm OST}(M_1,M_2)=\hbar\int d^3k\,\bigg{[}(M_1^2-M_2^2)[\omega_k^1\hat{a}_{1,\bar{k
}}a_{1,\bar{k}}+\omega_k^2\hat{a}_{2,\bar{k}}a_{2,
\bar{k}}]+\frac{1}{2}(\omega_k^1+\omega_k^2)\delta^3(0)\bigg{]},
\label{E70}
\end{equation}
where $\delta^3(0)=\int d^3x/(2\pi)^3$ is a momentum space delta function that serves as the spatial volume. In (\ref{E69}) we see that both of the $[a_{1,\bar{k}},\hat{a}_{1,\bar{k}^{\prime}}]$ and $[a_{2,\bar{k}},\hat{a}_{2,\bar{k}^{\prime}}]$ commutators are positive, and in (\ref{E70}) we see that there are no states with negative energy.  There are thus no states with negative norm or negative energy, and we thus achieve our primary objective of establishing that the second- plus fourth-order theory is fully consistent and unitary \cite{R21a}.

While we can also anticipate that the pure fourth-order theory of interest to us will be unitary too, explicitly showing it to be so requires care since (\ref{E69}), (\ref{E70}) and the partial fraction decomposition in (\ref{E61}) all be come ill-defined if let $M_1$ and $M_2$ go to zero. Because of this, we need to treat the limit theory separately, and will present a discussion of the needed limiting procedure in Sec. \ref{S8}.

However before doing so, at this point we can now revisit the Hamiltonian we obtained for 2D quantum Einstein gravity as given in (\ref{E22}), since we have now seen that $\cal{P}\cal{T}$-invariant Hamiltonians can have real eigenvalues even if they are not Hermitian, and in such cases fields can have expansions in which the creation and annihilation operators are not conjugates of each other. When we analyzed the 2D quantum Einstein theory, we had found that if we took the field operators to be Hermitian, the Hamiltonian in (\ref{E22}) was then not Hermitian. However, with all the solutions to the wave equations in (\ref{E7}) having real energies, we now see that 2D quantum Einstein gravity emerges as a $\cal{P}\cal{T}$ theory with a $\cal{P}\cal{T}$-invariant Hamiltonian. Consequently, the operators in the field expansion of (\ref{E15}) need to be given a structure similar to that displayed in (\ref{E68}). However since doing this does not affect the energies of the eigenmodes (the energy eigenvalues of the right eigenvectors $|R \rangle$ in (\ref{E65}) are not affected by the relation of the $\langle R|$ conjugates of the right eigenvectors to the left eigenvectors $\langle L|$), the cancellation of gravitational and matter field zero-point energy densities in 2D quantum Einstein gravity remains intact. Since 4D conformal gravity is also a $\cal{P}\cal{T}$ theory, we thus see that both of these conformal theories are in the $\cal{P}\cal{T}$ rather than the Hermitian class.

\subsection{The Ghost Problem of Pure Fourth-Order Theories}
\label{S8}

To explore the limit in which we switch off both $M_1$ and $M_2$, we note first that in this limit the two independent solutions to the wave equation (\ref{E60}) with a given $\bar{k}$, viz. the plane waves $\psi_1=e^{i(\bar{k}\cdot \bar{x} -\omega^1_kt)}$ and $\psi_2=e^{i(\bar{k}\cdot \bar{x} -\omega^2_kt)}$, behave in a very unusual way. Specifically, in the limit they both collapse onto one and the same wave function $\psi=e^{i(\bar{k}\cdot \bar{x} -\omega_kt)}$ where $\omega_k=|\bar{k}|$. However, since the wave equation remains fourth order, it cannot lose any of its solutions, and there thus has to be another solution to $(-\partial_t^2+\bar{\nabla}^2)^2\phi(x)=0$ with the same $\bar{k}$. To find it we expand $\psi_1$ and $\psi_2$ to first order in $M_1^2$ and $M_2^2$, to obtain
\begin{equation}
\psi_1 \rightarrow e^{i(\bar{k}\cdot \bar{x} -\omega_kt)}\left(1-\frac{iM_1^2t}{2\omega_k}\right),\qquad
\psi_2 \rightarrow e^{i(\bar{k}\cdot \bar{x} -\omega_kt)}\left(1-\frac{iM_2^2t}{2\omega_k}\right).
\label{E71}
\end{equation}
Thus by combining $\psi_1$ and $\psi_2$ with singular weights, we can construct a combination that is not singular in the limit, viz.
\begin{equation}
\psi_-=\frac{2i\omega_k}{(M_1^2-M_2^2)}\left(\psi_1 -\psi_2\right) \rightarrow e^{i(\bar{k}\cdot \bar{x} -\omega_kt)}t.
\label{E72}
\end{equation}
The missing solution to $(-\partial_t^2+\bar{\nabla}^2)^2\phi(x)=0$ is thus power-behaved in $t$. Comparing now with (\ref{E52}), we recognize the $(n \cdot x)=-t$ term in it as having none other than the form of the solution given in (\ref{E72}). The $(n \cdot x)=-t$ factor in (\ref{E52}) thus arises from the singular nature of the limiting procedure, and $n^{\mu}$ is a timelike vector since the limiting procedure only involves  the frequency-dependent terms in $\psi_1$ and $\psi_2$.

Despite the fact that the $\psi_-$ solution grows linearly in time, in this solution the classical $H^{\rm OST}(M_1=0,M_2=0)$ is still time independent, as it has to be time independent in any solution. Thus the structure of the classical $H^{\rm OST}(M_1=0,M_2=0)$ is such that runaways in time do not cause runaways in energy. However, in the quantum theory, unlike the $\psi_+=(\psi_1+\psi_2)/2 \rightarrow e^{i(\bar{k}\cdot \bar{x} -\omega_kt)}$ solution, the $\psi_-$ solution is not an eigenstate of $i\partial_t$. Consequently, in the quantum theory the quantum Hamiltonian has lost an eigenstate, with $\psi_-$ being found to be a solution to the time-dependent Schr\"odinger equation but not the time-independent one. Of the two solutions, one is stationary and the other is not, with the Hamiltonian losing eigenstates in the limit. Despite this, the set of stationary plus non-stationary solutions combined is still complete (since solutions to the time-independent Schr\"odinger equation are also solutions to the time-dependent one, the solutions to the time-dependent Schr\"odinger equation were already complete before the limit was taken), with the pure fourth-order theory still being unitary \cite{R2}.

Because the set of stationary solutions alone is not complete in the limit, the quantum  $H^{\rm OST}(M_1=0,M_2=0)$ does not possess a complete set of energy eigenstates. Even though the dimensionality of the full Hilbert space of the field operators is that of a two-dimensional oscillator, the dimensionality of the space of eigenstates of $H^{\rm OST}(M_1=0,M_2=0)$ is that of a one-dimensional one. Since $H^{\rm OST}(M_1=0,M_2=0)$ does not have a complete set of eigenstates, it cannot be diagonalized, and is thus in non-diagonalizable Jordan-block form. Since it cannot be diagonalized, it cannot be Hermitian, and since it is the limit of an $H^{\rm OST}(M_1,M_2)$ with non-zero $M_1$ and $M_2$, the quantum $H^{\rm OST}(M_1,M_2)$ could not have been Hermitian either, just as we had found. Moreover, since $H^{\rm OST}(M_1=0,M_2=0)$ cannot be diagonalized, the similarity transformation that brings $H^{\rm OST}(M_1,M_2)$ to a Hermitian form must be singular in the limit too, just as had been found \cite{R22}.

With regard to Jordan-block matrices, we recall that Jordan had shown that by a sequence of similarity transformations, any matrix can be brought to the Jordan canonical form in which it is composed of blocks that are either diagonal or in a non-diagonalizable triangular form in which all the elements on one side of the diagonal are zero.  A typical example of a non-diagonalizable Jordan-block matrix is the two-dimensional
\begin{eqnarray}
M=\pmatrix{1&1\cr 0&1}.
\label{E73}
\end{eqnarray}
Its secular equation $|M-\lambda I|=0$ has two solutions for $\lambda$, both of which are equal to one (and incidentally both real even though $M$ is not Hermitian), but  $M$ has only one right-eigenvector, and equally, only one left-eigenvector, viz.
\begin{eqnarray}
\pmatrix{1&1\cr 0&1}\pmatrix{1\cr 0}=\pmatrix{1\cr 0},\qquad \pmatrix{0 & 1}=\pmatrix{0 & 1}\pmatrix{1&1\cr 0&1}.
\label{E74}
\end{eqnarray}

To effect the $M_1^2\rightarrow 0$, $M_2^2 \rightarrow 0$ limit of the Hamiltonian, the field operators and the commutation relations of (\ref{E70}), (\ref{E68})  and (\ref{E69}), in analog to \cite{R2} it is very convenient to set 
\begin{eqnarray}
&&a_{\bar{k}}=\mu_{\bar{k}}(a_{1,\bar{k}}+ia_{2,\bar{k}})+ \frac{b_{\bar{k}}}{2},\qquad 
\hat{a}_{\bar{k}}=\mu_{\bar{k}}
(\hat{a}_{1,\bar{k}}+i\hat{a}_{2,\bar{k}})+\frac{\hat{b}_{\bar{k}}}{2},
\nonumber\\
&&b_{\bar{k}}=\lambda_{\bar{k}}(a_{1,\bar{k}}-ia_{2,\bar{k}}),\qquad \hat{b}_{\bar{k}}=\lambda_{\bar{k}}
(\hat{a}_{1,\bar{k}}-i\hat{a}_{2,\bar{k}}),
\nonumber \\
&& \mu_{\bar{k}}=\frac{(\omega^1_k+\omega^2_k)}{\surd{2}},\qquad  \lambda_{\bar{k}}=\frac{(\omega^1_k-\omega^2_k)}{\surd{2}},
\label{E75}
\end{eqnarray}
with the Hamiltonian, the field operators, and the commutation relations then limiting to 
 \begin{eqnarray}
H^{\rm OST}(M_1=0,M_2=0)&=&\hbar \int d^3k\,\omega_k[
\hat{a}_{\bar{k}}b_{\bar{k}}+a_{\bar{k}}\hat{b}_{\bar{k}}+2\hat{b}_{\bar{k}}b_{\bar{k}}],
\nonumber\\
&=&\hbar \int d^3k\,\omega_k[
\hat{a}_{\bar{k}}b_{\bar{k}}+\hat{b}_{\bar{k}}a_{\bar{k}}+2\hat{b}_{\bar{k}}b_{\bar{k}}+\delta^3(0)],
\label{E76}
\end{eqnarray}
 \begin{eqnarray}
\phi(x)&=&\int \frac{d^3k~\hbar^{1/2}}{(2\pi)^{3/2}2(\omega_k)^{3/2}}\bigg{[}e^{i(\bar{k}\cdot \bar{x}-\omega_kt)}\left[a_{\bar{k}} +i\omega_k(n \cdot x)b_{\bar{k}}\right]
+e^{-i(\bar{k}\cdot \bar{x}-\omega_kt)}\left[\hat{a}_{\bar{k}} -i\omega_k(n \cdot x)\hat{b}_{\bar{k}}\right]\bigg{]},
\nonumber \\
\pi^0&=&\partial_{\lambda}\partial^0\partial^{\lambda}\phi.\qquad \pi^{00}=-\partial^0\partial^0\phi,
\label{E77}
\end{eqnarray}
\begin{eqnarray}
&&[a_{\bar{k}},\hat{b}_{\bar{k}^{\prime}}]=[b_{\bar{k}},\hat{a}_{\bar{k}^{\prime
}}]=\delta^3(\bar{k}-\bar{k}^{\prime}),
\nonumber\\
&&[a_{\bar{k}},\hat{a}_{\bar{k}^{\prime}}]=0,\quad[b_{\bar{k}},\hat{b}_{\bar{k
}^{\prime}}]=0,\quad[a_{\bar{k}},b_{\bar{k}^{\prime}}]=0,\quad[\hat{a}_{\bar{k}}
,\hat{b}_{\bar{k}^{\prime}}]=0,
\label{E78}
\end{eqnarray}
where $\omega_k=|\bar{k}|$.
As constructed, $H^{\rm OST}(M_1=0,M_2=0)$ has a vacuum zero-point energy equal to $\hbar\omega_k$ for each 3-vector $\bar{k}$, just as befits a two oscillator theory, a value that is precisely double the zero-point energy of a single oscillator. In the one-particle sector there is only one eigenstate, namely $\hat{b}|\Omega\rangle$, with $\hat{a}|\Omega\rangle$ not being an eigenstate at all. Moreover, since the $[b_{\bar{k}},\hat{b}_{\bar{k}^{\prime}}]$ commutator is zero, the state $\hat{b}|\Omega\rangle$ has zero norm. We thus confirm that $H^{\rm OST}(M_1=0,M_2=0)$ is indeed a non-diagonalizable Jordan-block operator with an incomplete set of eigenstates, with the presence of zero-norm states also being a characteristic of non-diagonalizable Jordan-block matrices. (The overlap of the left- and right-eigenvectors of the typical Jordan-block matrix of (\ref{E73}) is expressly zero.) 

Having now constructed $H^{\rm OST}(M_1=0,M_2=0)$ in the form given in (\ref{E76}), we note that if we had not been aware that $H^{\rm OST}(M_1=0,M_2=0)$ is not Hermitian, we would have reached a contradiction. Specifically, suppose we had started directly with the $M_1=0$, $M_2=0$ theory with $I_{\rm S}=-(1/2)\int d^4x\partial_{\mu}\partial_{\nu}\phi\partial^{\mu}\partial^{\nu}\phi$, and initially taken $\phi(x)$ to be Hermitian. The same Ostogradski quantization procedure would have led us to the exact same structure exhibited in (\ref{E76}), (\ref{E77}) and (\ref{E78}), except that the hatted operators then would have been the Hermitian conjugates of the unhatted ones, and we would have obtained an $H^{\rm OST}(M_1=0,M_2=0)$ that appeared to be Hermitian. However, such an $H^{\rm OST}(M_1=0,M_2=0)$ would still be found to be missing some energy eigenstates and not be diagonalizable, and hence it could not possibly be Hermitian. Consequently, we would have had to have gone right back to the starting point and not taken $\phi(x)$ to be Hermitian at all. Thus in higher-order theories of gravity the quantum gravitational field cannot be Hermitian, with unitarity problems only having been encountered because one treated fields as though they were.

To now explicitly quantize the pure fourth-order gravity theory, on comparing (\ref{E58}) with (\ref{E76}), we recognize that the conformal gravity $-4\alpha_g\int d^3 x W_{00}(2)$ as constructed from the gravitational equations of motion has precisely the same structure as $H^{\rm OST}(M_1=0,M_2=0)$ as constructed via the Ostrogradski procedure \cite{R23}. However, before identifying them we need to comment on an oddity in the sign conventions of the metric-based definition of $T_{\mu\nu}$ and the canonical definition. To clarify the issue, consider the simple example of the scalar field action $I=\int d^4x (-g)^{1/2}{\cal L}$ where ${\cal L}=(1/2)\partial_{\mu}\phi\partial^{\mu}\phi$. For this action one canonically defines $\pi_{\mu}=\partial {\cal L}/\partial\phi_{,\mu}=\partial_{\mu}\phi$, and sets $T^{\mu\nu}_{\rm CAN}=\partial^{\mu}\phi\partial^{\nu}\phi-g^{\mu\nu}{\cal L}$. However, for the same action the metric-based definition evaluates to $T^{\mu\nu}_{\rm MET}=2(-g)^{-1/2}\delta I/\delta g_{\mu\nu}=-\partial^{\mu}\phi\partial^{\nu}\phi+g^{\mu\nu}{\cal L}$, i.e. to the opposite sign. To identify the origin of the difference we note that the functional variation of $(-g)^{1/2}$ is given by $\delta (-g)^{1/2}=(1/2)(-g)^{1/2}g^{\mu\nu}\delta g_{\mu\nu}$, to thus give $T^{\mu\nu}_{\rm MET}\sim +g^{\mu\nu}{\cal L}$. However, for $T^{\mu\nu}_{\rm CAN}$, one wants $T^{\mu\nu}_{\rm CAN}$ to behave as the Legendre transform of the Lagrangian. For a metric with signature $g_{00}=+1$, for the scalar field action we find that $T^{00}_{\rm CAN}$ is given by the positive definite $T^{00}_{\rm CAN}=\dot{\phi}^2-{\cal L}=(1/2)[\dot{\phi}^2+\bar{\nabla}{\phi}^2]$. Similarly,  for a metric with signature $g_{00}=-1$, to get the same positive value for $T^{00}_{\rm CAN}$ one has to set ${\cal L}=-(1/2)\partial_{\mu}\phi\partial^{\mu}\phi$. The difference in overall sign between $T^{\mu\nu}_{\rm CAN}$ and $T^{\mu\nu}_{\rm MET}$ is not a metric signature issue or a choice in the overall sign of ${\cal L}$, but is an intrinsic overall sign difference that exists between the $T^{\mu\nu}_{\rm CAN}$ and $T^{\mu\nu}_{\rm MET}$ for any given choice of sign of $g_{00}$ and ${\cal L}$ provided one uses the same choice of signs in the two cases \cite{R24}.

For the fourth-order derivative case of interest to us here, on comparing the transverse gauge conformal gravity action of (\ref{E51}), viz.  $I_{\rm W}=-(\alpha_g/2)\int d^4x \partial_{\alpha}\partial^{\alpha} K_{\mu\nu}\partial_{\beta}\partial^{\beta} K^{\mu\nu}$ with  the massless limit of the scalar action of (\ref{E59}), viz. $-(1/2)\int d^4x \partial_{\alpha}\partial^{\alpha}\phi\partial_{\beta}\partial^{\beta} \phi$, we see that the metric based $T^{\mu\nu}_{\rm MET}$ associated with $I_{\rm W}$ and the canonical  $T^{\mu\nu}_{\rm CAN}$ associated with $I_{\rm S}$ will differ in overall sign if we take $I_{\rm W}$ and $I_{\rm S}$ to be equal, i.e. if we set $\alpha_g=1$. To accommodate this difference we need to treat $\alpha_g$ as though it had been multiplied by an additional minus sign. Thus, on allowing for the explicit factor of $(-\alpha_g)^{1/2}$ in the definition of $K_{\mu\nu}$ given in (\ref{E53}), we see that the conformal gravity Hamiltonian given in (\ref{E58}) will be positive definite if we identify the operators in (\ref{E58}) and (\ref{E53}) with those in (\ref{E76}) and (\ref{E77}) according to $\hat{B}^{(i)}(\bar{k})\sim \hat{b}_{\bar{k}}$, $B^{(i)}(\bar{k})\sim b_{\bar{k}}$, $\hat{A}^{(i)}(\bar{k})\sim \hat{a}_{\bar{k}}$, $A^{(i)}(\bar{k})\sim a_{\bar{k}}$. From (\ref{E78}),  the quantization of gravitational $K_{\mu\nu}(x)$ fields given in (\ref{E53}) thus has to be of the form
\begin{eqnarray}
&&[A^{(i)}(\bar{k}),\hat{B}^{(j)}(\bar{k}^{\prime})]=[B^{(i)}(\bar{k}),\hat{A}^{(j)}(\bar{k}^{\prime})]=\delta_{i,j}\delta^3(\bar{k}-\bar{k}^{\prime}),
\nonumber\\
&&[A^{(i)}(\bar{k}),\hat{A}^{(j)}(\bar{k}^{\prime})]=0,\qquad [B^{(i)}(\bar{k}),\hat{B}^{(j)}(\bar{k}^{\prime})]=0,
\nonumber\\
&&[A^{(i)}(\bar{k}),B^{(j)}(\bar{k}^{\prime})]=0,\qquad [\hat{A}^{(i)}(\bar{k}),\hat{B}^{(j)}(\bar{k}^{\prime})]=0,
\label{E79}
\end{eqnarray}
with $-4\alpha_g\int d^3 x W_{00}(2)$ then taking the form
\begin{eqnarray}
&&-4\alpha_g\int d^3 x  W_{00}(2) 
\nonumber\\
&&\qquad =\sum_i\hbar\int d^3k\,\omega_k\left[\hat{A}^{(i)}(\bar{k})B^{(i)}(\bar{k})+\hat{B}^{(i)}(\bar{k})A^{(i)}(\bar{k})+2\hat{B}^{(i)}(\bar{k})B^{(i)}(\bar{k})+\delta^3(0)\right].
\label{E80}
\end{eqnarray}

Given the structure of (\ref{E79}) and (\ref{E80}), certain comments are in order. First we see that the $[B^{(i)}(\bar{k}),\hat{B}^{(j)}(\bar{k}^{\prime})]$ commutator vanishes in (\ref{E79}), and we thus precisely recover the relations in (\ref{E56}) that we had imposed earlier.  As regards the particle content of the theory, we note that of the two one-particle states, one of them, $\hat{A}^{(i)}(\bar{k})|\Omega \rangle$, is not an eigenstate at all, while the other, $\hat{B}^{(i)}(\bar{k})|\Omega \rangle$, is a state of zero norm, a state that can thus leave no imprint in a detector. In the conformal theory then there is no observable, on-shell, positive norm graviton at all \cite{R25}. While this is at variance with the way one ordinarily thinks of gravitons, namely as quantized gravity waves, once one takes gravity to be intrinsically quantum-mechanical, there is then no classical gravity wave to quantize in the first place. Any observable gravity wave or c-number gravitational field present in the theory would have to be associated with a matrix element of the quantum gravitational field in a state with an indefinite number of gravitons, in much the same manner as a classical electromagnetic wave is constructed in QED. \cite{R26}. That the structure of the conformal gravity graviton is so much at variance with the standard Einstein gravity view of gravitons, is because in the conformal theory the gravitational field does not exist as an a priori classical field. Because of this, the quantum gravitational field need not have a conventional classical limit, and in fact does not, with the gravitational field not being Hermitian and the gravitational Hamiltonian not only not being Hermitian either, it is not even diagonalizable,

The fact that the $\hat{A}^{(i)}(\bar{k})|\Omega \rangle$ state is not an eigenstate in the quantum theory is a reflection of the zero-energy theorem \cite{R27} of classical conformal gravity. Specifically, Boulware, Horowitz, and Strominger studied modes in the classical conformal theory that have the same asymptotic properties as the modes of the standard second-order Einstein theory, and found them to have zero energy, and thus not propagate. (An analogous result can be found in \cite{R28}.) Since in the expansion of $K_{\mu\nu}$ given in (\ref{E53}), it is the $A^{(i)}(\bar{k})$ and $\hat{A}^{(i)}(\bar{k})$ modes that are akin to standard gravity modes, and since it is the non-asymptotically flat $B^{(i)}(\bar{k})$ and $\hat{B}^{(i)}(\bar{k})$ modes with their linear in $t$ dependence that are not, the zero-energy theorem is to apply to the $A^{(i)}(\bar{k})$ and $\hat{A}^{(i)}(\bar{k})$ modes alone in the limit in which the $B^{(i)}(\bar{k})$ and $\hat{B}^{(i)}(\bar{k})$ modes are excluded. Examining now $-4\alpha_g\int d^3x W_{00}(2)$ as given in (\ref{E58}) \cite{R29}, we see that because there are no $A^{(i)}(\bar{k})\hat{A}^{(i)}(\bar{k})$ cross-terms, on setting the $B^{(i)}(\bar{k})$ and $\hat{B}^{(i)}(\bar{k})$ terms to zero, when treated classically the quantity $-4\alpha_g\int d^3x W_{00}(2)$ becomes zero, to thus give the zero-energy theorem. In the quantum theory, this same absence of $A^{(i)}(\bar{k})\hat{A}^{(i)}(\bar{k})$ cross-terms entails that there are no associated one-particle eigenstates. The zero-energy theorem of classical conformal gravity thus translates into the need for the Hamiltonian to be Jordan-block in the quantum theory. However, the zero-energy theorem should not be thought of as saying that all the modes of the theory have zero energy as the theorem does not apply to the non-asymptotically flat $B^{(i)}(\bar{k})$ and  $\hat{B}^{(i)}(\bar{k})$ modes. For these modes there are appropriate cross-terms, and these modes can have, and indeed do have, non-zero energy.

Even though there is a zero-energy theorem for the $A^{(i)}(\bar{k})$ and $\hat{A}^{(i)}(\bar{k})$ modes in the classical theory, inspection of (\ref{E80}) shows that in the quantum theory these modes do have some effect on the energy, as they do contribute to the zero-point energy that is obtained when the $A^{(i)}(\bar{k})\hat{B}^{(i)}(\bar{k})$ product is replaced by the $\hat{B}^{(i)}(\bar{k})A^{(i)}(\bar{k})$ product through the use of the $[A^{(i)}(\bar{k}),\hat{B}^{(i)}(\bar{k})]$ commutator. As regards the zero-point energy in the graviton sector, from (\ref{E80}) we find it to be equal to $\hbar\omega_k$ for each $\bar{k}$ and each polarization state $\epsilon^{(i)}(\bar{k})$, i.e. an $\hbar\omega_k$ for each fourth-order (i.e. double oscillator) polarization state. Thus even though the $\hat{A}^{(i)}(\bar{k})|\Omega \rangle$ are not one-particle eigenstates and cannot materialize on shell as observable particles, they still contribute to the zero-point energy. As such, this is reminiscent of the 2D quantum Einstein gravity case where, even though there was no graviton propagation, there was still a zero-point contribution. In conformal gravity then, the $\hat{A}^{(i)}(\bar{k})|\Omega \rangle$ modes behave more like the graviton of 2D Einstein gravity than the graviton of the 4D Einstein theory.

With the net zero-point energy of a conformal gravity polarization state being $\hbar\omega_k$ for each $\bar{k}$, since there are two polarization states in the gravitational field basis, the total contribution to the zero-point energy for a given $\bar{k}$ is $2\hbar\omega_k$, a quantity, which when integrated over all momenta gives the quartically divergent $\hbar K^4/4\pi^2$. Now we recall that the zero-point energy of a free massless 2-component 4D fermion is $-\hbar\omega_k$, to thus give $-2\hbar\omega_k$ for each $\bar{k}$ of a 4-component one. In 4D conformal gravity then, the quartically divergent zero-point energies of the graviton and a 4-component Dirac fermion precisely cancel each other. Thus when we couple conformal gravity to a free massless 4-component fermion, the quantization of the fermion field  forces the gravitational field to be quantized too, and gives the gravitational field the exact zero-point energy needed to cancel that which the fermion obtained from its own quantization. In this way then, massless 4D gravitons and fermions precisely solve each other's zero-point energy problem, just as desired.

\section{Conclusions and Comments}
\label{S9}

In this paper we have provided a comprehensive treatment of the cosmological constant, zero-point energy, and quantum gravity problems. The discussion presented here immediately raises several further issues that require investigation. As regards the 4D zero-point cancellation, one has to ask exactly how it is to be achieved if there are more fields than just the one 4-component fermion whose zero-point energy exactly  cancels that of the gravitational field. One has to ask how the cancellation is to be maintained in the presence of dynamical symmetry breaking. One has to ask how one is able to explain the fact that cosmological observations seem to require a cosmological constant that is not cancelled completely, and one needs to ask what are the specific observational predictions of the theory that would allow one to test it. 

If one extends the matter sector to $M$ massless gauge bosons and $N$ massless two-component fermions, the net quartic divergence that they generate will be due to $M-N$ units of $\hbar\omega_k$ for each $\bar{k}$. (For gauge bosons one gets $+\hbar\omega_k/2$ for each of two helicity states.) Since the vanishing of the total $T^{\mu\nu}_{\rm UNIV}=T^{\mu\nu}_{\rm GRAV}+T^{\mu\nu}_{\rm M}$ of (\ref{E24}) is a mathematical identity, the gravitational field sector is required to generate a net $Z$ times $2\hbar\omega_k$ where $2Z+M-N=0$ \cite{R29a}. However, without any internal symmetry, one does not have $Z$ gravitons, and thus the consistency of (\ref{E24}) forces the graviton to not be quantized canonically, but to have commutators in (\ref{E79}) that are normalized to $Z\delta^3(\bar{k}-\bar{k}^{\prime})$ rather than to $\delta^3(\bar{k}-\bar{k}^{\prime})$ itself. 

To clarify the nature of this requirement, we note  when one quantizes a theory canonically, the canonically constructed energy-momentum tensor has the same normalization as the canonically constructed canonical conjugates, with $H=\int d^3x T^{00}$ being the time translation generator that enforces $[\phi,H]=i\hbar \dot{\phi}$. With all normalizations being fixed by the canonical quantization prescription, one cannot obtain $[\phi,H]=i\hbar Z\dot{\phi}$ instead. However, while we do use canonical quantization for the matter fields, we do not use it for the gravitational field, as it is to be quantized by virtue of its being coupled to the matter fields. Now, in a linearization around flat spacetime, one still needs to be able to define a Hamiltonian that obeys $[\phi,H]=i\hbar \dot{\phi}$ for the gravitational field components as flat spacetime is Poincare invariant. To construct such a Hamiltonian we note that since variation of the gravitational action with respect to the metric produces a $T^{\mu\nu}_{\rm GRAV}$ that  is covariantly conserved, the quantity $\int d^3x T^{00}_{\rm GRAV}(2)$ has the transformation properties that a linearized gravitational Hamiltonian is required to have  \cite{R30}, and will continue to have them even if it is multiplied by a constant. If we have a wave equation with frequencies $\omega_k$ and a quantization procedure that gives a generic Hamiltonian of the form $\sum (\hbar \omega_k/2)(a^{\dagger}_ka_k +a_ka_k^{\dagger})$, then with a quantization rule of the form $[a_k,a^{\dagger}_k]=Z$, the one-particle state $a^{\dagger}_k|\Omega\rangle$ would have energy $Z\hbar \omega_k$ rather than the needed $\hbar \omega_k$. Thus with a quantization rule of the form $[a_k,a^{\dagger}_k]=Z$, the Hamiltonian would need to be given by $\sum (\hbar \omega_k/2)(a^{\dagger}_ka_k +a_ka_k^{\dagger})/Z=\sum \hbar \omega_k(a^{\dagger}_ka_k/Z+1/2)$ instead. And while one cannot make such a modification for the matter fields, one can do so for the gravitational field as the relation of its time translation generator to the spatial integral of its $T^{00}_{\rm GRAV}(2)$ is not specified by the structure of the gravitational sector itself. Rather, it is specified only after one couples to the matter fields, i.e. only after it is forced to be quantized in the first place. Thus the coupling of gravity to a source containing more than one matter field forces  the quantity $\int d^3x T^{00}_{\rm GRAV}(2)/Z$ to be the time translation generator for the gravitational field, being so even as  $\langle \Omega|\int d^3x T^{00}_{\rm GRAV}(2)|\Omega \rangle$ contributes $2Z\hbar \omega_k$ per $\bar{k}$ to $\langle \Omega|\int d^3x T^{00}_{\rm UNIV}|\Omega \rangle$ \cite{R31}.

With the normalization constant $Z$ needing to obey $Z=(N-M)/2$, the positivity of $Z$ imposes some constraints on model building \cite{R32}. For the standard $SU(3)\times SU(2)\times U(1)$ theory for instance, we have $M=12$ gauge bosons and $N=16$ two-component spinors per generation, with $Z$ then being positive. For the grand-unified gauge group $SO(10)$ one has $M=45$ and again $N=16$ per generation, with three generations of fermions being the minimum number that would make $Z$ be positive this time. If one wishes to put all of the generations of fermions into a single irreducible representation of a grand-unifying group, for an $SU(n)$ group with the gauge bosons in the adjoint and the fermions all in one fundamental there are no $N-M>0$ solutions at all; while for an $SO(2n)$ group with all the fermions in one spinor representation, solutions are obtained for $2n \geq 16$, with the smallest possibility being an $SO(16)$ with eight fermion generations where $M=120$ and $N=128$.  The $SO(16)$ values for $M$ and $N$ readily extend to the exceptional group $E_8$ since its adjoint decomposes into  $248=120+128$ under its $SO(16)$ subgroup \cite{R32a}. Included within the set of $2n \geq 16$ solutions is the $SO(18)$ grand-unifying group considered in \cite{R32b}.  Moreover, groups such as $SO(16+2k)$ contain $SO(10)\times SO(6+2k)$ as a subgroup, with $SO(6+2k)$ linking the $16$-dimensional spinor representations of $SO(10)$ much as for instance described in \cite{R32c}. 

The emergence of such $SO(2n \geq 16)$ grand-unifying groups via zero-point considerations is welcome since such groups are triangle anomaly free. Thus even without requiring anomaly cancellation per se, we are still led to grand-unifying groups in which the cancellation occurs. Additional constraints can be imposed if one also requires that the gauge boson/fermion sector be asymptotically free. As noted in \cite{R32b}, $SO(2n)$ groups with all the fermions in one irreducible spinor representation of the group are only asymptotically free up to $SO(20)$ since beyond that there are too many fermions. With zero-point cancellation leading to $SO(2n \geq 16)$, we see that out of all possible $SU(n)$ and $SO(2n)$ grand-unifying groups that one might consider, only $SO(16)$, $SO(18)$ and $SO(20)$ meet all the constraints. That this range is so narrow is because the positivity of $Z=(N-M)/2$ favors fermions over bosons, while the negativity of the gauge theory renormalization group beta function favors bosons over fermions, to only leave a narrow window in which both sets of constraints can be met.

With regard to the issue of mass generation by dynamical symmetry breaking, we note that as long as our two key stationarity and trace conditions, $T^{\mu\nu}_{\rm UNIV}=T^{\mu\nu}_{\rm GRAV}+T^{\mu\nu}_{\rm M}=0$ and $g_{\mu\nu}T^{\mu\nu}_{\rm M}=0$ continue to hold, the cancellations found in the 2D case will continue to occur after a fermion bilinear condensate acquires a non-vanishing vacuum expectation value. However, as with the 2D case, the gravitational field commutator $Z$ factor will acquire a dependence on the dynamically induced mass parameters. And again, all zero-point and induced cosmological constant terms will mutually cancel each other, provided only that both the gravity and matter field sectors are conformal and renormalizable and gravity is purely quantum-mechanical. On the gravity side this means conformal gravity and on the matter side it means the standard fermion and gauge boson gauge theories but with no fundamental Higgs fields with their non-conformal double-well potentials.

While the needed cancellations are actually guaranteed to occur once all the above conditions are met, constructing an explicit  model in which all the cancellations are manifestly seen to occur is technically very difficult. One cannot simply work with the standard Nambu-Jona-Lasinio model, since in 4D, the model is neither conformal invariant nor renormalizable, and the trace of the matter field energy-momentum tensor is non-zero. One thus has to look at dynamical symmetry breaking by fermion condensates in a 4D theory that is both conformal invariant and renormalizable. We thus need to have dynamical symmetry breaking occur in a theory of massless fermions and gauge bosons. In such theories, radiative corrections will lead to Callan-Symanzik scaling anomalies, and will destroy the conformal invariance of the theory in the ultraviolet (though not in the infrared where dynamical symmetry breaking takes place). However, if the theory is at a renormalization group fixed point, then as noted by Wilson, the conformal invariance will be restored, only with anomalous rather than canonical dimensions for the field operators. When anomalous dimensions turn out to be lower than their canonical values, the short-distance behavior of the theory will be softened, a property that enabled Johnson, Baker and Willey \cite{R33} to construct a theory of QED in which the renormalization constants were then finite. 

As the ultraviolet behavior becomes less divergent, at the same time the behavior of the theory in the infrared becomes more divergent. However, unlike ultraviolet divergences, infrared divergences are actually welcome in a sense, since they can lead to spontaneous symmetry breaking and long range order. We thus take note of the study of the Johnson-Baker-Willey theory that was given in \cite{R34}. In \cite{R34} it was found that if the dimension $d_{\theta}=3+\gamma_{\theta}$ of the fermion composite bilinear $\theta=\bar{\psi}\psi$ is reduced by one whole unit from its canonical value of three to an anomalous value of two so that the insertion of $\bar{\psi}\psi$ into the inverse fermion propagator would behave as $\tilde{\Gamma}_{\theta}(p,p,0)=(-p^2/M^2)^{-1/2}$, the vacuum would then undergo dynamical symmetry breaking and generate a fermion mass. In this case the mean-field Nambu-Jona-Lasinio type zero-point energy density would change from $\epsilon(m)=(i/\hbar)\int d^4p/(2\pi)^4{\rm Tr Ln}(\gamma^{\mu}p_{\mu}-m+i\epsilon)$ to $\epsilon(m)=(i/\hbar)\int d^4p/(2\pi)^4{\rm Tr Ln}(\gamma^{\mu}p_{\mu}-m(-p^2/M^2)^{-1/2}+i\epsilon)$. With $M^4=16\hbar^4K^4{\rm exp}(8\pi^2\hbar^3/M^2g)$, $\epsilon (m)-m^2/2g$ then evaluates to 
\begin{equation}
\epsilon(m)-\frac{m^2}{2g}=-\frac{\hbar K^4}{4\pi^2}+ \frac{m^2M^2}{16\pi^2\hbar^3}\left[{\rm ln}\left(\frac{m^2}{M^2}\right)-1\right], 
\label{E81}
\end{equation}
to thus have a completely finite $m$-dependent term with minima at $m=\pm M$. Deeply reducing the dynamical dimension of the very same composite operators that are to cause dynamical symmetry breaking can thus force the expectation values of these operators to actually be non-zero. Thus at this $\gamma_{\theta}=-1$ critical value, the infrared divergences of the massless theory oblige the expectation value of  $\bar{\psi}\psi$ to move away from zero, with the mass then being generated self-consistently. While the zero-point energy density of a non-interacting 4D fermion of mass $m$ would contain mass-dependent quadratic and logarithmic divergences in addition to the mass-independent quartic divergence $-\hbar K^4/4\pi^2$ that it already has when it is massless, in an interacting theory at the critical value of $\gamma_{\theta}=-1$ the mass-dependent divergence is found \cite{R34} to be reduced to logarithmic only (i.e. two whole units of reduction in $\bar{\psi}\psi \bar{\psi}\psi$ for each reduction of one unit in $\bar{\psi}\psi$, with $\bar{\psi}\psi \bar{\psi}\psi$ now having dynamical dimension equal to four, to thus be non-perturbatively renormalizable). The logarithmic divergence is then cancelled identically by a logarithmically divergent induced cosmological constant term $\Lambda_{\rm MF}=-m^2/2g$, just as happened in (\ref{E35}) for the 2D mean-feld $T^{\mu\nu}_{\rm MF}$. In (\ref{E81}) the energy density  is thus left with a mass-independent quartic divergence and a mass-dependent finite part that evaluates to $-M^4/16\pi^2\hbar^3$ at the minimum. Because of (\ref{E24}), these two terms must both be cancelled by the graviton zero-point energy density. Thus, at $\gamma_{\theta}=-1$ the 4D cancellation completely parallels the cancellation of the quadratically divergent and finite terms (c.f. (\ref{E35}) and (\ref{E36})) that was found above in the 2D conformal case \cite{R35}. 

While the analysis of QED at $\gamma_{\theta}=-1$ is an attempt to implement the ides of Nambu and Jona-Lasinio within a renormalizable context, there is one key distinction. Specifically, despite the fact that there is dynamical mass generation in the theory, there is no Goldstone boson \cite{R33}.  As noted by Baker and Johnson, in scale-invariant theories there is an anomalous evasion of the Goldstone theorem. What happens is that the bare fermion mass $m_0$ is not zero identically (as it would be in the Nambu and Jona-Lasinio case). Rather it only goes to zero in the limit of infinite cut-off (as $K^{\gamma_{\theta}}$ where $\gamma_{\theta}$ is negative). At the same time the renormalization constant $Z_{\theta}$ associated with the operator $\theta=\bar{\psi}\psi$ goes to infinity as $K^{-\gamma_{\theta}}$, with the product $m_0\bar{\psi}\psi$ being finite. In consequence of this anomaly, there is no pole in the fermion anti-fermion scattering amplitude even though the mass non-trivially obeys a self-consistent gap type equation \cite{R36}. In a conformal gravity theory that is realized with anomalous dimensions, one might expect an analog of the Baker-Johnson evasion to occur.

In the study presented in this paper we have concentrated on the properties of the vacuum. However, some interesting changes occur once one starts populating the positive energy fermion and boson states as well. Specifically, one can then find coherent states $|C\rangle$ in which $m(x)=\langle C|\bar{\psi}(x)\psi(x)|C\rangle$ becomes a spacetime-dependent Ginzburg-Landau order parameter.  (For a discussion with references to the literature see \cite{R37,R38}.) Two types of spacetime dependencies are of particular interest, a time-dependent one for cosmology and a space-dependent one for macroscopic systems such as stars and for microscopic systems such as elementary particles. With the space-dependent case being discussed in \cite{R37,R38} and references therein, we comment here only on the time-dependent situation. When matrix elements of the energy-momentum tensor are evaluated in coherent states, in addition to all the vacuum terms we have encountered above, one also get derivatives of the order parameter as well. Such coherent states can be associated \cite{R38} with the stationary variation of the vacuum functional $-\int d^4x W(m(x))=-\int d^4x(-\epsilon(m(x))+(1/2)Z(m(x))\partial^{\mu}m(x)\partial_{\mu}m(x)+....)$ where $\epsilon(m(x))$ is the vacuum energy as evaluated in the state with $m=m(x)$. At the stationary extremum, the equation of motion for $m(x)$ that ensues will look just like that of a Ginzburg-Landau effective theory, with the  $Z(m(x))$ term making a kinetic energy type contribution. Then, since the total $T^{\mu\nu}_{\rm UNIV}$ of (\ref{E24}) vanishes, its matrix elements in coherent states will vanish too, to still give a grand cancellation. However, now the cancellation will have to include the kinetic energy contribution as well. Since this contribution is absent in the vacuum, and since the trace of the matter field $T^{\mu\nu}_{\rm M}$ still vanishes, the kinetic energy term will be equal to the amount by which the quantity $p_{\rm MF}-\rho_{\rm MF}-4\Lambda_{\rm MF} $ (viz. (\ref{E33}) as written in 4D) changes as one evaluates the mean-field fermion energy-momentum tensor in states $|C\rangle$ rather than in states $|S\rangle$. It is the time dependence of $m(t)$ that is recognized as the Robertson-Walker scale parameter in an expanding cosmology, and it  will be coupled to the residual change in $p_{\rm MF}-\rho_{\rm MF}-4\Lambda_{\rm MF} $ rather than to these quantities themselves. The cosmological constant can thus be huge and yet its effect on cosmic evolution would still be small (symbolically behaving as $\langle C|\bar{\psi}\psi|C\rangle$-$\langle S|\bar{\psi}\psi|S\rangle$), with the coherent state in which it is to be evaluated redshifting as the universe expands. 

In \cite{R4} cosmology was discussed within the framework of a particularly chosen Ginzburg-Landau effective order parameter theory, and a very good fit to the accelerating universe Hubble plot data was obtained.  The challenge posed by the work of this paper then is to see what effective Ginzburg-Landau theory it leads to for cosmology, and what departures from homogeneity and isotropy it then  produces in the cosmic microwave background.

In our work we have  required symmetry breaking to be dynamical, with the scalar Higgs field that is commonly used in symmetry breaking in particle theory having to only be a c-number order parameter in an effective Ginzburg-Landau theory. Being a c-number, such a Higgs field would not be detectable as a particle in an accelerator experiment, and in addition, there would be no quadratically divergent self-energy hierarchy problem. Moreover, there would be no need for any fundamental double-well Higgs potential with its tachyonic mass term, a potential that if it exists only serves to exacerbate the cosmological constant problem, since there appears to be nothing that would specify where the zero of the potential is to be located. However, while there should be no fundamental Higgs field, if the symmetry is broken in a theory in which bare fermion masses are zero identically, there would instead be a dynamically generated massive scalar bound state in the fermion anti-fermion scattering amplitude \cite{R3}. Unfortunately, little is known as to the value of its mass, or into which specific channels it is to decay, with the current experimental bounds obtained from fundamental Higgs particle searches being of little guidance. Finally, in theories in which bare fermion masses are not zero identically, but only go to zero as the cut-off goes to infinity, there would still be an effective Ginzburg-Landau order parameter theory (as explicitly constructed in the third reference in \cite{R34}), but because of the Baker-Johnson evasion, there would be no massless bound states at all. Now it was noted in \cite{R37,R38} that when fermions that undergo dynamical symmetry breaking are coupled to external gauge fields, because of the underlying gauge invariance of the fermion and gauge boson couplings, in the effective Ginzburg-Landau Lagrangian that is induced the order parameter will minimally couple to the gauge field. In the presence of a non-vanishing order parameter the gauge field wave equation would then describe a massive gauge field. As noted in \cite{R38}, if this same effect were to occur in a theory where there is simultaneously a Baker-Johnson evasion of the Goldstone theorem,  there would then be massive gauge fields and no observable Higgs particles at all.

To conclude this paper, we would like to make some comments on what one should expect of a quantum gravity theory. We begin by observing that a straightforward reading of gravitational equations of motion such as the Einstein equations $-(1/8\pi G)G^{\mu\nu}=T^{\mu\nu}$ would equate classical terms on either side to each other and equate quantum field-theoretic terms on either side to each other since the equation of motion is an operator identity. However, for practical applications of Einstein gravity, it is assumed that the Einstein equations can be reinterpreted as relating a classical $G^{\mu\nu}$ to a c-number matrix element of the quantum-mechanical components of $T^{\mu\nu}$. To derive such a semi-classical approximation, one has to be able to dominate the quantum-mechanical path integral by a stationary phase in which the gravitational field strength is large. However, there appears to be no established justification for this approximation, as the terms that one ignores are not negligible. Rather, they are actually infinite because of the lack of renormalizability of 4D quantum Einstein gravity, The objective of theories such as string theory is to use string properties to cancel all these undesirable divergences and recover the semi-classical Einstein equations. An acceptable quantum theory of gravity then is one in which one can derive semi-classical equations to use for gravitational phenomenology \cite{R39}. Now the issue of deriving a semi-classical limit from a quantum theory is also met in electrodynamics, and there a stationary phase approximation is reliable because quantum electrodynamics is renormalizable. Thus it is natural to try to do the same thing for gravity, and one is thus led to consider conformal gravity as one is able to interpret its equation of motion  $ 4\alpha_g W^{\mu\nu}=T^{\mu\nu}_{\rm M}$ as a bona fide quantum operator relation whose associated path integral is well-behaved in four dimensions \cite{R40}. And as we have seen, such an approach leads to a resolution of the zero-point and cosmological constant problems that has yet to be achieved in theories such as string theory \cite{R41}. As a final comment on our work we note that by requiring that curvature be entirely due to quantum effects, we not only change the way that one ordinarily thinks about gravity, we essentially eliminate one of the central challenges that one faces in constructing a quantum gravity theory starting from a classical one. Specifically, with there then being no a priori classical curvature, one does not have to make it compatible with quantization.

{}

\begin{thebibliography}{}
\bibitem{R5} To keep track of all the relevant $\hbar$ factors associated with quantization, throughout this paper we define $k$ and $\omega_k$ in wave numbers rather than momenta. While not crucial to our work here, we note in passing that in a sense it is more natural to use wave numbers rather than momenta since field theory is based on waves rather than particles. These waves obey wave equations that are based on wave numbers rather than momenta, with particles and their momenta only emerging following a quantization of the fields. Moreover, giving precedence to wave numbers over momenta allows a reformulation of the black-body formula that enables us to address a peculiarity of it. Specifically, for a black-body photon gas we recall that the energy density is given by $U/V=\pi^2(k_{\rm B}T)^4/15(\hbar c)^3$ where $k_{\rm B}$ is Boltzmann's constant. As constructed, we see that as we let $\hbar \rightarrow 0$ this expression goes to infinity. Now while there is no classical black-body to which a quantum black-body could non-trivially be required to limit to, it would still be more natural if $U/V$ were to go to zero rather than infinity in the $\hbar \rightarrow 0$ limit, as quantum effects ordinarily disappear in the classical limit. As a formula, the black-body formula itself is nothing more than a sum over photon one-particle energies $\hbar \omega$ as weighted by a temperature-dependent Bose-Einstein occupation number factor. Since there is no classical one-particle photon, again it would be more natural to have $U/V$ go to zero rather than to infinity in the limit. To understand why $U/V$ does diverge as $\hbar \rightarrow 0$, we recall that the occupation number weighting factor behaves as $\langle n_k\rangle=2/(e^{\hbar \omega/k_{\rm B}T}-1)$, from which it follows that $U=\sum \hbar \omega_k \langle n_k\rangle$ blows up as $1/\hbar^3$. However, suppose we instead work in wave numbers and define a 'wave number' temperature $\hat{T}=T/\hbar$. In this case $\langle n_k\rangle$ would behave as $2/(e^{\omega/k_{\rm B}\hat{T}}-1)$, and $U/V$ would behave as $U/V=\pi^2\hbar(k_{\rm B}\hat{T})^4/15c^3$ and then be linear in $\hbar$, just like the one-particle energy itself. In this sense then it is perhaps more natural to use wave numbers, with one then being able to consider both mass (which can be introduced via a 'wave number' inverse Compton wavelength mass according to $\omega_k=(k^2+m^2/\hbar^2)^{1/2}$) and temperature as being intrinsically quantum-mechanical. With the black-body finite temperature one-particle energy density being written as $U/V=\pi^2\hbar(k_{\rm B}\hat{T})^4/15c^3$ and with the photon zero-point energy density being given as $\sum \hbar\omega/2$, both of these energy density terms are in the same linear in $\hbar$ form that we use for the fermion zero-point energy density throughout this paper.
\bibitem{R6} In (\ref{E15}) we take the $A(k)$, $B(k)$, $C(k)$ and $D(k)$ operators to have the same $1/k^{1/2}$ dimension as standard 2D boson creation and annihilation operators. While we do not quantize the gravitational field canonically, we note that for a non-total divergence action of  generic form $(1/\kappa_2^2)\int d^2x\partial_0h_{00}\partial^0h^{00}$, the canonical conjugate of $h^{00}$ would be given by $(1/\kappa_2^2)\partial_0h_{00}$ and the canonical commutator would be of form $[h^{00},(1/\kappa_2^2)\partial_0h_{00}]=i\hbar\delta(x-x^{\prime})$. To parallel this structure, in the field expansion of $h_{\mu\nu}$ in (\ref{E15}) we therefore include an overall factor $\kappa_2\hbar^{1/2}$.
\bibitem{R12} In this respect our approach differs from the situation that occurs when classical gravity  is coupled to a classical source. Specifically, in the purely classical case a classical background is perturbed by a first order classical perturbation $\delta T^{\mu\nu}_{\rm M}$, with this perturbation then serving as a source to the first order metric fluctuation in an equation such as $ -(1/\kappa_4^2)G^{\mu\nu}(1)=\delta T^{\mu\nu}_{\rm M}$ in the Einstein case, or $4\alpha_g W^{\mu\nu}(1)=\delta T^{\mu\nu}_{\rm M}$ in the conformal case. Since first order equations of motion such as these cannot be satisfied trivially once $\delta T^{\mu\nu}_{\rm M}$ is non-zero, we see that in the classical case already in first order the fluctuation $h_{\mu\nu}$ is forced to be non-zero. Moreover, in the classical case $h_{\mu\nu}$ does not satisfy a free wave equation such as $W^{\mu\nu}(1)=0$ or $G^{\mu\nu}(1)$. Rather, it satisfies a wave equation with a source term. If the $\delta T^{\mu\nu}_{\rm M}$ source just happens to be localized in space, then far from the source the fluctuation will obey a free wave equation, though otherwise not. However, in the strictly quantum situation there is no first order $\delta T^{\mu\nu}_{\rm M}$, and the first order fluctuation will satisfy a free wave equation identically on all distance scales no matter what the spatial dependence of the quantum-mechanical $T^{\mu\nu}_{\rm M}$ might be.
\bibitem{R13} In passing we note that in standard 4D Einstein gravity, the fluctuation $h_{\mu\nu}$ has ten components, and thus can only be reduced to six by gauge transformations, five transverse ones and the trace $h=\eta_{\mu\nu}h^{\mu\nu}$ itself. One cannot reduce $h_{\mu\nu}$ to five components via gauge transformations, and in standard gravity once has to keep all six components. Now while there is equally no way to reduce $h_{\mu\nu}$ to five components in conformal gravity either, nonetheless, one does not have to take the trace of $h_{\mu\nu}$ into consideration, since in conformal gravity the trace plays no role. Thus in practice, in conformal gravity one only has to deal with five components.
\bibitem{R19} While the viewpoint advocated in this paper is that gravity is quantized because of its coupling to quantized matter, for pure 4D conformal gravity it turns out that the quantum structure that ensues when gravity is coupled to a non-interacting massless 4D fermion is the same as the one obtained by a canonical quantization procedure (essentially because in  4D conformal gravity there are no ordering problems of the type met in the 2D conformal case), and so for simplicity we follow the canonical procedure here. However, as already noted in our discussion of 2D quantum Einstein gravity, once the fermion gets a dynamical mass, there are changes in the commutation structure on the gravitational side, changes that cannot be obtained via a canonical quantization procedure.
\bibitem{R22} If we drop the spatial dependence of the fields in $T^{\rm OST}_{00}$ so that $\omega_1=M_1$, $\omega_2=M_2$, $T^{\rm OST}_{00}$ reduces to the Hamiltonian $H_{\rm PU}$ of the Pais-Uhlenbeck oscillator model that was given in \cite{R21}, viz. $T^{\rm OST}_{00}\rightarrow H_{\rm PU}=\pi_{0}\dot{\phi}+\frac{1}{2}\pi_{00}^2+\frac{1}{2}(\omega_1^2+\omega_2^2)\dot{\phi}^2-\frac{1}{2}\omega_1^2\omega_2^2\phi^2$. For its energy levels to be bounded from below, $H_{\rm PU}$ cannot  be Hermitian (if $\phi$ were Hermitian the $-(1/2)\omega_1^2\omega_2^2\phi^2$ term would be unbounded from below). And in \cite{R1,R2} it was shown that when the energies are bounded from below and $\phi$ and $H_{\rm PU}$ are not Hermitian, $H_{\rm PU}$ can be transformed to a Hermitian form via $\tilde{H}_{\rm PU}= e^{-{\cal{Q}}/2}H_{\rm PU}e^{{\cal{Q}}/2}$ where ${\cal{Q}}=i\alpha \pi_{00} \pi_0-i\beta \phi\dot{\phi}$, $\alpha=(1/\omega_1\omega_2){\rm ln}[(\omega_1+\omega_2)/(\omega_1-\omega_2)]$, and $\beta=\alpha \omega_1^2\omega_2^2$. In the equal frequency limit this transformation is singular and $H_{\rm PU}$ can no longer be diagonalized.
\bibitem{R24} The issue of the difference between $T^{\mu\nu}_{\rm MET}$ and $T^{\mu\nu}_{\rm CAN}$ would also be met for the fermionic field. However in Secs. \ref{S2} and \ref{S3} we never introduced a canonical fermionic $T^{\mu\nu}_{\rm CAN}$ at all, with  the $T^{\mu\nu}_{\rm MET}$ that we did introduce being negative definite in the vacuum, just as it should be.
\bibitem{R26} While the issue of the existence or not of gravity waves in the conformal theory remains to be explored, we note that this has no direct bearing on the observed decay of the orbit of a binary pulsar. Specifically, one can describe the binary pulsar effect entirely in terms of near-zone gravitational radiation reaction, without needing to treat the effect as a far-zone radiation emission effect at all. Since the conformal gravity $1/k^4$ propagator has the standard retardation properties of covariant theories with gravitational information being transmitted at a finite speed, virtual graviton exchange between the two stars can cause the orbit to decay even if no graviton could ever become real and go on shell in the far zone.
\bibitem{R29} While we have derived (\ref{E58}) here by taking it to be quantum-mechanical, since there are no ordering problems in 4D conformal gravity, one would obtain exactly the same generic expression if one did treat $W_{00}(2)$ classically, with there being no $A^{(i)}(\bar{k})\hat{A}^{(i)}(\bar{k})$ cross-terms in a classical treatment either.
\bibitem{R29a} Since all bosons (including the graviton) have positive zero point energy density and all fermions have negative zero-point energy density (in any spacetime dimension),  one could never get a total zero-point energy density cancellation from bosonic particles alone. The very consistency of gravity theory thus requires the existence of fermions.
\bibitem{R31} Even though it is not necessary for the matter fields, in passing we note that one could also quantize them starting from the metric-derived $T^{\mu\nu}_{\rm M}$ rather than from the canonically-derived  energy-momentum tensor, and then determine the quantization rules from the requirement that $H=\int d^3x T^{00}_{\rm M}$ generate $[\phi,H]=i\hbar \dot{\phi}$ for the matter fields. Since one can construct $T^{\mu\nu}_{\rm M}$ directly via metric variation of the matter action without reference to matter field constraints, one might be able to bypass the constraint problem (or perhaps some aspects of it) that is met when one tries to treat constrained matter field systems canonically.
\bibitem{R32} If $Z$ is negative, one would be back to the original Pais and Uhlenbeck realization of fourth-order theories in which one avoids negative norms by instead having the energy be unbounded below, with a now Hermitian gravitational field then providing negative zero-point energy to counterbalance the positive net zero-point energy in the matter sector. However, it is not clear if one can give a sensible physical interpretation to this realization, since as noted in \cite{R2}, it would require evaluating (\ref{E61}) using an unusual contour in which some positive energy modes would propagate  backward in time and some negative energy modes would propagate forward.
\bibitem{R35} It is instructive to compare the role of the cosmological constant in the dynamical and fundamental Higgs field symmetry breaking cases. For the double-well Higgs potential the value of the potential at its minimum contributes to the cosmological constant term (term $g^{\mu\nu}V(\phi)$ in $T^{\mu\nu}$), with the minimization of the $\Lambda$-dependent potential leading to a large, but finite, cosmological constant term. In the dynamical case the role of the cosmological constant term is to cancel divergences in the trace of the energy-momentum tensor, with the finite piece that is to be minimized being entirely in the zero-point energy density.  Thus for $T_{\rm MF}^{\mu\nu}=(\rho_{\rm MF}+p_{\rm MF})U^{\mu}U^{\nu}+p_{\rm MF}g^{\mu\nu}-\Lambda_{\rm MF} g^{\mu\nu}$, the trace condition $3p_{\rm MF}-\rho_{\rm MF}-4\Lambda_{\rm MF}=0$ allows us to write $T_{\rm MF}^{\mu\nu}=(\rho_{\rm MF}+p_{\rm MF})[U^{\mu}U^{\nu}+g^{\mu\nu}/4]$, with its $\Lambda_{\rm MF}$-independent finite part being minimized in (\ref{E81}).
\bibitem{R36} If there is a dynamical Goldstone pole in the fermion anti-fermion scattering amplitude, its residue will obey a self-consistent equation. However, in and of itself, the fact that there might be a non-trivial solution to this self-consistent equation does not mean that there necessarily is a massless pole. To establish the presence of a Goldstone pole one has to go back to the fermion anti-fermion scattering amplitude and look at it at energies away from $q^2=0$ to see whether or not a pole sets in as one moves the energy to $q^2=0$ itself. When this is done in the Johnson-Baker-Willey case, no pole is found.
\bibitem{R39} For geodesic behavior all that is required is that one can obtain a semi-classical approximation in which the matter field is coupled in its wave equation to a c-number gravitational field matrix element, with eikonalized short wavelength rays then being geodesic. It is not necessary that the Christoffel symbol terms that appear in the geodesic equation be independent of $\hbar$. They only need to be c-numbers. Thus if we consider cosmology in coherent states $|C\rangle$, the Robertson-Walker metric that will be generated will possess a c-number scale factor $a(t)$ and a c-number spatial 3-curvature $k$ that will both depend on $\hbar$. However, for phenomenological Hubble plot fitting for instance one only needs to know the time dependence of $a(t)$ and the sign of $k$ if it is non-zero, and one can absorb any factors of $\hbar$ into the overall coefficients and proceed as though one is doing a purely classical calculation. (This of course is exactly how one does treat standard cosmology in the case where the source is a black-body, viz. a system whose matrix elements explicitly depend on $\hbar$.)
\bibitem{R40} While beyond the scope of the present paper, with conformal gravity being a  consistent quantum theory, it might be able to address the problem of the collapse of the wave function in measurement theory, with the collapse perhaps only being an apparent one in which one neglects the quantum gravitational component of the wave function.
\bibitem{R41} While the supersymmetry of string theory leads to an exact cancellation of fermion and boson zero-point energies and requires a zero cosmological constant term before the supersymmetry is broken, unlike the conformal gravity case, supersymmetry has no control over these quantities once the symmetry  is broken and the superpartners acquire unequal masses. In string theory not only is it proving very challenging to produce a small cosmological constant at all, at the very same time one would have to produce TeV region supersymmetric superparticles so as to account for their lack of detection to date. However, if one were to endow conformal gravity with a supersymmetry so as to  enlarge it to a conformal supergravity theory, all of the required cancellations would still go through no matter how large the superscale might be.
\end{thebibliography}
\end{document}